\documentclass[a4paper,fleqn,usenatbib]{mnras}

\usepackage{newtxtext,newtxmath}

\usepackage[T1]{fontenc}
\usepackage{ae,aecompl}

\usepackage{graphicx}	
\usepackage{amsmath}	
\usepackage{amssymb}	
\usepackage{epsfig}
\usepackage{animate}

\title[AGN loci in BPT]{Upper boundaries of AGN regions in optical diagnostic diagrams}

\author[X. Ji et al.]{Xihan Ji$^{1}$\thanks{Contact e-mail: \href{mailto:xji243@uky.edu}{xji243@uky.edu}}, Renbin Yan$^{1}$\thanks{Contact e-mail: \href{mailto:rya225@g.uky.edu}{rya225@g.uky.edu}},
Rog\'erio Riffel$^{2,3}$,
Niv Drory$^{4}$, and
Kai Zhang$^{5}$
\\
$^{1}$Department of Physics and Astronomy, University of Kentucky, 505 Rose Street, Lexington, KY 40506, USA\\
$^{2}$Departamento de Astronomia, Universidade Federal do Rio Grande do Sul - Av. Bento Gon\c calves 9500, Porto Alegre, RS, Brazil.
\\
$^{3}$Laborat\'orio Interinstitucional de e-Astronomia, Rua General Jos\'e Cristino, 77 Vasco da Gama, Rio de Janeiro, 20921-400, Brazil
\\
$^{4}$McDonald Observatory, The University of Texas at Austin, 1 University Station, Austin, TX 78712, USA
\\
$^{5}$Lawrence Berkeley National Laboratory, 1 Cyclotron Road, Berkeley, CA 94720, USA}

\date{Accepted XXX. Received YYY; in original form ZZZ}

\pubyear{2020}

\begin{document}
\label{firstpage}
\pagerange{\pageref{firstpage}--\pageref{lastpage}}
\maketitle

\begin{abstract}
The distribution of galaxies in optical diagnostic diagrams can provide information about their physical parameters when compared with ionization models under proper assumptions. By using a sample of central emitting regions from the MaNGA survey, we find evidence of the existence of upper boundaries for narrow-line regions (NLRs) of active galactic nuclei (AGN) in optical BPT diagrams, especially in the diagrams involving [S~II]$\lambda \lambda$6716, 6731/H$\alpha$. Photoionization models can well reproduce the boundaries as a consequence of the decrease of [S~II]$\lambda \lambda$6716, 6731/H$\alpha$ and [O~III]$\lambda$5007/H$\beta$ ratios at very high metallicity. Whilst the exact location of the upper boundary in the [S~II] BPT diagram only weakly depends on the electron density of the ionized cloud and the secondary nitrogen prescription, its dependence on the shapes of the input SEDs is much stronger. This allows us to constrain the power-law index of the AGN SED between 1 Ryd and $\sim100$ Ryd to be less than or equal to $-1.40\pm 0.05$. The coverage of the photoionization models in the [N~II] BPT diagram has a stronger dependence on the electron density and the secondary nitrogen prescription. With the density constrained by the [S~II] doublet ratio and the input SED constrained by the [S~II] BPT diagram, we find that the extent of the data in the [N~II] BPT diagram favors those prescriptions with high N/O ratios. Although shock-ionized cloud can produce similar line ratios as those by photoionization, the resulting shapes of the upper boundaries, if exist, would likely be different from those of a photoionizing origin.
\end{abstract}




\section{Introduction}

Optical diagnostic diagrams (hereafter BPT) are useful tools to discriminate different photoionization sources for galaxies when optical spectra are available (\citealp{1981PASP...93....5B, 1987ApJS...63..295V}). A typical BPT diagnostic sorts galaxies into three categories: H~II regions, composite regions, and AGNs (for the [N~II]-based BPT diagnostic, cf.~\citealp{2001ApJ...556..121K}, \citealp{2003MNRAS.346.1055K}), or star-forming (SF) regions, low-ionization nuclear emission-line regions (LINERs, cf.~\citealp{1980A&A....87..152H}) and Seyferts (for the [S~II]- or [O~I]-based BPT diagnostic, cf.~\citealp{2006MNRAS.372..961K}). Different parts of galaxies occupy specific regions on the diagrams, with H~II and SF regions lying on the lower left, Seyferts taking up the space on the upper right, LINERs occupying the lower right and composite regions sitting in the middle. By comparing observational data with the predictions from photoionization models, people have established a theoretical basis for the existence of the SF loci in optical diagnostic diagrams \citep[e.g.][]{1981PASP...93....5B, 1987ApJS...63..295V, 2000ApJ...542..224D, 2001ApJ...556..121K}. This provides a way to put constraints on physical parameters like gas-phase metallicity or ionization parameter for SF regions (\citealp{2000ApJ...542..224D, 2002ApJS..142...35K}). An interesting feature of the SF regions is that it has an upper boundary on all BPT diagrams. This upper boundary was both confirmed observationally by \cite{2003MNRAS.346.1055K} with SDSS data, and theoretically by \cite{2001ApJ...556..121K} with modeling of extreme starburst galaxies. The commonly accepted explanation for the existence of the boundary is that the drop of electron temperature becomes more important at higher metallicity, which will subsequently lower the strength of collisional excited lines like [N~II]$\lambda \lambda$6548, 6583 and [S~II]$\lambda \lambda$6716, 6731.

For other regions on the BPT diagrams, however, less is known about how their physical parameters are connected with their positions on the planes. One of the reasons is that ionization sources other than photons from young OB stars play an important role in these regions. The physics of such regions is not yet fully understood. Active galactic nuclei (AGN) with different ionizing powers have been considered as the primary driven engines for Seyferts and LINERs (\citealp{2006MNRAS.372..961K, 2008ARA&A..46..475H}). However it has been shown by many studies that LINER-like spectra do not always indicate a nuclear origin (\citealp{2006MNRAS.366.1151S, 2010MNRAS.403.1036C, 2011MNRAS.413.1687C, 2012ApJ...747...61Y, 2013A&A...558A..43S, 2016MNRAS.461.3111B, 2017ApJ...851L..24H}). Post-AGB stars (\citealp{1994A&A...292...13B}) as well as shocks (\citealp{1995ApJ...455..468D}) and turbulent mixings (\citealp{1993ApJ...407...83S}) might play an important role in creating LINER-like spectra. They could all contribute in some degree, while the weight of each component might well change for different galaxies or different locations inside one galaxy. Therefore in the rest of the paper we call LINER-like regions LI(N)ERs. Despite all these difficulties, it is of great significance to establish a match between the theoretical models and positions of Seyferts and LI(N)ERs in BPT diagrams. This will result in easily accessible and reliable calibrators for chemical abundance and other important parameters for these regions (\citealp{1998AJ....115..909S}). This approach will inherit the advantages of BPT diagnostics: only involving strong optical lines and not sensitive to dust extinction, which is usually another parameter difficult to estimate for these regions.

There have been various attempts to understand the AGN region with modelling of ionized gas. The simplest and most direct approach is to assume a photoionization dominated scenario. Other ionizing sources can be evaluated later as independent components. Many models have been provided to reproduce the line-ratio distribution of the narrow-line regions (NLRs), among which the dusty, ionization pressure dominated photoionization model seems promising and is in good agreement with the observations (\citealp{2004ApJS..153....9G, 2014MNRAS.437.2376R, 2016MNRAS.456.3354F}). Under this assumption, it is natural to expect that AGN regions may also have upper boundaries as H~II regions do in normal BPT diagrams. And \cite{2004ApJS..153...75G} did find that their models would result in degeneracy at both high metallicity and high ionization parameter. Though it remains a question whether the observed AGNs ever reach such regime of the parameter space. The AGN boundary, if exists, could serve as a starting baseline to calibrate the ionized gas in the two dimensional parameter space of gas-phase metallicity and ionization parameter, provided we understand how other physical parameters might influence its position on the diagnostic plane. Conversely one can use the observational boundary of a sample of AGN ionized clouds to interpret the properties of underlying AGNs by matching this boundary with the theoretical ones. But so far no detailed discussion has been made about how this boundary can come into being and what physical interpretation can be made. In this paper we try to demonstrate the following points: the observational data indicate the existence of upper-right boundaries for AGN regions in the [S~II] and [N~II] BPT diagrams; the boundaries match quite well the predictions of photoionization models; we can use this boundary to put constraints on some physical parameters of AGN regions. The paper is organized as the following. In Section 2 we describe the observational data. In Section 3 we show our sample selection criteria and sample distribution. In Section 4 we present the photoionization models we use and discuss the physical meanings of the AGN boundary. In section 5 we briefly discuss shock models as a possible contributor and the difficulties with the [O~I] BPT diagram. We summarize our results and present our conclusions in Section 6.

Throughout this paper, we assume a $\Lambda$CDM model with H$_0 = 70$\,km\,s$^{-1}$Mpc$^{-1}$, $\Omega \rm _m=0.3$ and $\Omega _{\Lambda} =0.7$. All magnitudes are given in the AB system.

\section{Observational data}

We draw our sample from the eighth internal data release of Mapping Nearby Galaxies at Apache Point Observatory survey (MaNGA; \citealp{2015ApJ...798....7B, 2016AJ....152..197Y}), the MaNGA Product Launch 8 (MPL-8), which includes observations of 6507 galaxies. MaNGA is part of the Sloan Digital Sky Survey IV (\citealp{2017AJ....154...28B}) and is by far the largest integral field spectroscopy (IFS) survey, aiming at obtaining spatially resolved spectral information of $\sim$ 10000 galaxies with <z> $\sim$ 0.03 by 2020 (\citealp{2017AJ....154...86W}). MaNGA uses the 2.5 m Sloan Telescope (\citealp{2006AJ....131.2332G}) and Baryon Oscillation Spectroscopy Survey (BOSS) Spectrographs (\citealp{2013AJ....146...32S}) for observation. The light is fed to the dual-beam BOSS spectrographs through fiber bundles. The resulting wavelength coverage is from 3600 \AA\ to 10300 \AA, with a median spectral resulution R $\sim$ 2000. To achieve uniform spatial coverage for various galaxies, MaNGA adopts hexagonal fiber bundles with five different sizes, with the long-axis diameter ranging from 12$^{\prime \prime}$ to 32$^{\prime \prime}$, corresponding to 19-, 37-, 61-, 91-, and 127-fiber units,  respectively (\citealp{2015AJ....149...77D}). Flux calibration is done using another set of mini-bundles (\citealp{2016AJ....151....8Y}). The raw data are reconstructed into 3D data cubes by the Data Reduction Pipeline (DRP; \citealp{2016AJ....152...83L}), and the final spaxel size is $\sim$ 0$^{\prime \prime}_.$5 $\times$ 0$^{\prime \prime}_.$5. By contrast, the median FWHM of MaNGA's point spread function (PSF) is $\sim$ 2$^{\prime \prime}_.$5 (\citealp{2015AJ....150...19L}).

A data analysis pipeline (DAP; \citealp{2019AJ....158..231W, 2019AJ....158..160B}) has been developed to automatically process the reduced MaNGA spectra, providing maps of spatially resolved physical quantities. DAP utilizes pPXF (\citealp{2004PASP..116..138C, 2017MNRAS.466..798C}) to perform the fitting of stellar continuum and the emission lines in a muti-stage procedure. First, the stellar kinematics is determined using Voronoi binned data (\citealp{2003MNRAS.342..345C}). Then, with the kinematics fixed, we fit the spectrum in each spaxel to a combination of the stellar-continuum templates and the emission-line templates. The velocities of all emission lines are tied together while the velocity dispersions of different lines are allowed to vary independtly, except for the pair of lines in each doublet.

\section{Boundaries of AGN regions}

In this work we adopt the DAP products for our analysis. Our primary goal is to obtain a representative sample of spaxels from emission-line galaxy nuclei to study the distribution of AGN-powered line ratios. The selection procedure should be independent of the BPT diagnostics and should avoid contaminations from ionization sources unrelated with an AGN. Thus, we apply the following selection criteria.
\begin{enumerate}
    \item We select only spaxels from the nuclear region to reduce contamination from non-AGN-powered ionizng sources. Quantitatively we require R / R$\rm _e$ < 0.3, where R$\rm _e$ is the elliptical Petrosian effective semi-major axis of the galaxy measured in the SDSS r band from the NASA-Sloan Atlas \footnote{http://www.nsatlas.org}\citep[NSA;][]{Blanton-11}. For different galaxies in our sample, this corresponds to 0.4 $-$ 4.0 kpc. We also try generating a sample using a cut based on the FWHM of the MaNGA PSF in the SDSS r band, requiring R < FWHM / 2. This sample yields the same result and most importantly the same boundary, despite that the sample size is nearly halved. Therefore, we choose the first cut in this paper in order to present a more statistically significant result.
    \item We require good emission line detections and select only spaxels with signal-to-noise ratio (SNR) greater than 3 in all of the followng lines [S~II]$\lambda \lambda$6716, 6731, [N~II]$\lambda$6583, [O~III]$\lambda$5007, H$\alpha$, and H$\beta$. 
    \item We select only spaxels with strong emssion lines to reduce the contamination from LI(N)ERs with non-AGN origins. We adopt the value of 3 \AA\ for the equivalent width of H$\alpha$ line as a lower limit (\citealp{2011MNRAS.413.1687C, 2016MNRAS.461.3111B}). \cite{2019MNRAS.486.5075D} suggests a more stringent cut of 5 \AA . We have repeated our analysis using the more stringent threshold and found that the conclusions do not change.
    \item Spaxels should have relatively old stellar population to minimize the contamination of ionization by hot young stars. We require all spaxels in our sample to have D$\rm _n$(4000) > 1.8, where D$\rm _n$(4000) is defined by \cite{1999ApJ...527...54B} as the ratio of the average flux densities in two narrow continuum bands: 3850 - 3950 and 4000 - 4100 \AA, integrated following the formula given by \cite{1983ApJ...273..105B}. Using the stellar population synthesis model provided by \cite{2003MNRAS.344.1000B} (hereafter BC03), it is found that for a single-burst star formation history with solar metallicity described by BC03, this cut selects ages older than 3 Gyr (\citealp{2003MNRAS.341...33K}). This criterion will inevitably remove some true AGNs which are mixed with noticeable star formations. But we want our sample to be as pure as possible to locate the pure AGN sequence. We do find that a number of Seyferts are removed by the cut, which will be discussed in Section 4.4.
\end{enumerate}

\begin{figure*}
	\includegraphics[width=0.9\textwidth]{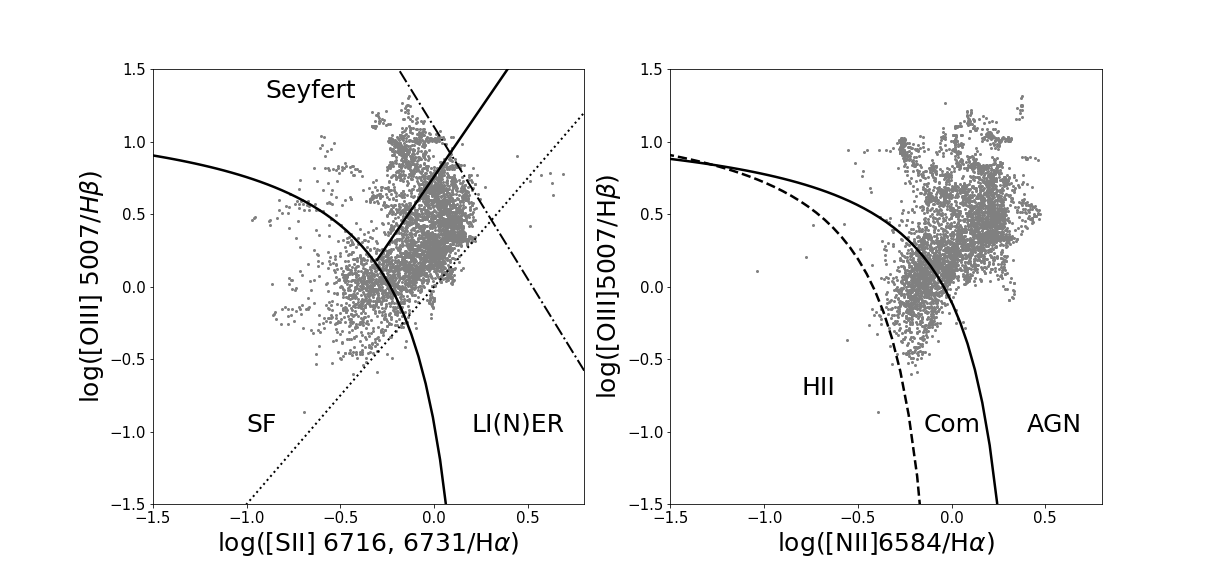}
	\caption{Distribution of sample spaxels on optical BPT diagrams. Left panel: [S~II]-based BPT diagram. The two demarcation lines are Kewley extreme starburst line (Ke01) which separate star-forming regions from AGN regions and \protect\cite{2006MNRAS.372..961K} line (Ke06) which separate Seyfert regions from LI(N)ER regions. Right panel: [N~II]-based BPT diagram. The dashed line is the empirical \protect\cite{2003MNRAS.346.1055K} line (Ka03) which separate H~II regions from composite regions. And the solid line is the Kewley extreme starburst line. The upper-right and lower-right boundaries we identify in the [S~II] BPT diagram are outlined by the dot-dashed line and the dotted line, respectively. There is no clear boundary in the [N~II] BPT diagram.}
    \label{fig:sample_dis}
\end{figure*}

Using these criteria, we end up with 4378 spaxels from 262 different galaxies, within which 122 galaxies contributes more than 10 spaxels each. Fig.~\ref{fig:sample_dis} shows the distribution of our sample spaxels on both [S~II]- and [N~II]-based BPT diagrams. In the [S~II] BPT diagram, the spaxels occupy all three regions with most of them classified as LI(N)ERs. There are two boundaries visible on the plane: one lies in the upper right while the other lies in the lower right. By eye we can pick two straight lines in Fig.~\ref{fig:sample_dis} which trace these boundaries. They can be written analytically as the following. For the upper right boundary, we have
\begin{equation}
    \text{log([O~III] 5007/H}\beta \text{)} = - 2.1 \text{log([S~II] 6716, 6731/H}\alpha \text{)} + 1.1 ,
    \label{eq1}
\end{equation}
and for the lower right boundary:
\begin{equation}
    \text{log([O~III] 5007/H}\beta \text{)} = 1.5 \text{log([S~II] 6716, 6731/H}\alpha \text{)} .
\end{equation}
Eq. ~\ref{eq1} has a good agreement with what we find later with photoionization models, where the intercept lies in a range from 0.90 to 1.00. The small difference from the observed boundary could be due to measurement uncertainties. 

In the [N~II] BPT diagram, the majority of our sample spaxels are classified as AGN or Composite regions. Unlike the case in the [S~II] BPT diagram, the distribution of the points at the upper right of the plane has more scattering. Hardly can one pick out a sharp boundary there.

\begin{figure}
	\includegraphics[width=0.5\textwidth]{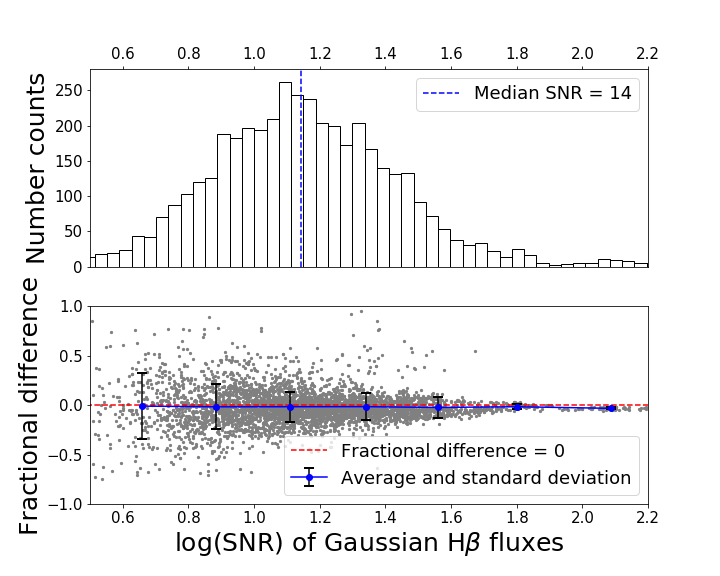}
	\caption{Data quality of the H$\beta$ line. Upper panel: Distribution of signal to noise ratios of the H$\beta$ line in our sample. The median value is 14. Lower panel: Fractional difference between the Gaussian flux and the summed flux of the H$\beta$ line (which is defined as (F$_{\rm Gaussian}$ $-$ F$_{\rm summed}$)/F$_{\rm Gaussian}$) versus the SNR. Error bars indicate the standard deviation for each bin.}
    \label{fig:measurement0}
\end{figure}

\begin{figure}
	\includegraphics[width=0.5\textwidth]{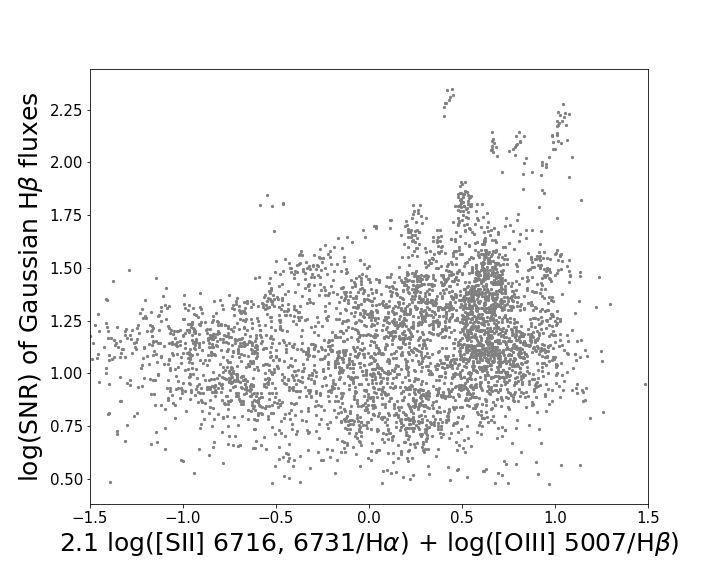}
	\caption{SNR of the H$\beta$ line versus the `distance' to the upper right boundary. The horizontal axis indicates the distance to a line parallel to the upper boundary in the [S~II] BPT diagram. The upper boundary has a value of 2.1 log([S~II] 6716, 6731/H$\alpha$) + log([O~III] 5007/H$\beta$) $\sim$ 1.1.}
    \label{fig:measurement1}
\end{figure}

Before we start to analyze the distribution of our sample, we would like to demonstrate that the obvious upper right boundary in the [S~II] BPT diagram is real, rather than due to poor data quality or selection effects. One might expect if the measurement of H$\beta$ line becomes less and less reliable towards the upper-right direction, an apparent boundary could emerge due to our cut in the SNR. Fig.~\ref{fig:measurement0} shows the overall data quality of the measurement of the H$\beta$ line in our sample. The median SNR of the H$\beta$ line is 14. The 16th-percentile and the 84th-percentile of the distribution of fractional differences between the summed fluxes and the Gaussian fluxes are $-15\%$ and 11\%, respectively. Basically, most of spaxels have strong lines that are way above the selection cuts. We also investigate the behavior of the SNR of H$\beta$ line as one moves towards this upper boundary. Fig.~\ref{fig:measurement1} shows that there is no obvious decreasing trend for the SNR of H$\beta$ line towards the upper boundary in the [S~II] BPT diagram, which further rules out the possibility of the boundary being caused by selection effects. The H$\beta$ line is supposed to be the weakest emission line among the lines we use, especially for those points near the boundary. As a sanity check, we also perform a similar analysis on [O~III]$\lambda$5007. In this case, the SNR is neither deceasing nor staying constant, but increasing towards the boundary. The summed fluxes and Gaussian fluxes have even better agreement. These results further ensure that the measurement of emission line ratios are accurate near the boundary. Note that in the [S~II] BPT diagram, there are a few spaxels that surpass both boundaries and lie far right on the plane. The emission lines of these spaxels have obvious broad-line components and are not properly fitted by the DAP. Therefore we discard these spaxels in the following analyses. Aware of the potential influence from the broadline components to the fitting of the narrow emission lines for the rest of our sample, we compared the fitted Gaussian flux of the [N II]-H$\alpha$ triplet to the directly summed flux of the three lines. The results show good consistency, which means that our single-component fitting can well recover the total flux in the great majority of cases.

We will demonstrate that AGN photoionization models provide a good explanation for the data distribution in BPT diagrams in Section 4. And in Section 5 we will explore the possibility of shock-ionization models.

\section{Photoionization models}

We use {\tt CLOUDY} (cf.~\citealp{2017RMxAA..53..385F}) to generate a series of photoionzation models for our analyses. There are several common assumptions for these models: 1. Constant gas pressure throughout the cloud; 2. Dust grains with typical ISM abundance (which will scale with the gas-phase metallicity of the cloud); 3. Metal depletion onto dust grains based on the values from \cite{1987ASSL..134..533J} and \cite{1986ARA&A..24..499C}; 4. Cosmic ray background of the local Universe. Note that because of the existence of metal depletion, the final gas-phase metallicity would be different from the initial one.

For each model, we vary the gas phase metallicity (oxygen abundance) from subsolar ($-$ 0.75 dex, or 0.178 Z$_{\odot}$, where the solar values are taken from \citealp{2010Ap&SS.328..179G}) to supersolar value (+ 0.75 dex, or 5.62 Z$_{\odot}$). All metals except carbon and nitrogen simply scale with the oxygen abundance. Nitrogen has a secondary origin and thus has a more complicated prescription. 
By default we use the nitrogen abundance prescription adopted by \cite{2004ApJS..153....9G}, which comes from fitting a compilation of H~II regions and nuclear starburst galaxies from \cite{2002A&A...389..106M} and \cite{2003ApJ...591..801K}. This presciption can be written as
\begin{equation}
    \rm N/O = 10^{-1.6} + 10^{2.33 + log(O/H)} .
    \label{eq_gro}
\end{equation}
The carbon abundance is set to be always 0.6 dex larger than the nitrogen abundance, as in the case of the solar abundance. Besides, we also consider the primary production of helium, which follows the equation in \cite{2002ApJ...572..753D}:
\begin{equation}
    \text{He / H} = 0.0737 + 0.024\cdot \text{Z / Z}\odot .
\end{equation}
We note that there is no common agreement on the choice of the nitrogen prescription. By the end of this section we will compare different formulations of the secondary elements and their results. Otherwise we will assume the above equations when setting up the models.

For the ionization parameter, which is defined as q $\equiv$ log($\rm \Phi _{\rm ion}$/n$\rm _{H}$c), where $\rm \Phi _{\rm ion}$ is the flux of photons with energy greater than 1 Ryd at the illuminated face of the cloud, we vary the value from - 4.5 to - 2.0. Observationally, the NLRs of Seyferts show small variation in line ratios such as [N~II]$\lambda \lambda$6548, 6583, [S~II]$\lambda \lambda$6716, 6731, and [O~III]$\lambda$5007/H$\beta$ (\citealp{2000A&ARv..10...81V}). And as we will see, a natural interpretation of the photoionzation models is that the q parameter of Seyferts rarely exceeds $-$ 2.0. However, \cite{2004ApJS..153...75G} argues that this could be an illusion due to the stagnation of line ratios at high q. We will go back to this issue later in this section.

Previous studies show that both metallicity and ionization parameter are unlikely to change much within 1 R$\rm _e$ of a single galaxy in the nearby universe (\citealp{2014A&A...563A..49S, 2017MNRAS.469..151B, 2018MNRAS.479.5235P}). So one would expect different galaxies occupy different regions within the model grids. As an example, a typical low-luminosity AGN like a LINER would have a metallicity above solar and a photoionization parameter of $\sim -3.5$ (\citealp{1983ApJ...264..105F, 1993ApJ...417...63H}). We can see that our sample spaxels have a relatively broad distribution in the BPT diagrams. This could be a result of varying intrinsic metallicities and ionization parameters. Alternatively, it can be explained by mixings of different ionization sources, i.e. an AGN-SF composite. We will check the plausibility of this scenario later.

\begin{table*}
	\centering
	\caption{Photoionization model set}
	\label{tab:model_sum}
	\begin{tabular}{l c}
		\hline
		\hline
		Parameter & Values \\
		\hline
		q & $-$ 4.5, $-$ 4.0, $-$ 3.5, $-$ 3.0, $-$ 2.5, $-$ 2.0 (also $-$ 1.5, $-$ 1.0, $-$ 0.5, 0.0 for some models)\\
		log (Z / Z$_\odot$) & $-$ 0.75, $-$ 0.50, $-$ 0.25, 0.0, 0.25, 0.50, 0.75\\
		log (n$\rm _e$ / cm$^{-3}$) & 1.0, 2.0, 3.0, 4.0\\
		Ionizing SED & Composite SED, Power-law SED ($\alpha$ = $-$ 1.2, $-$ 1.4, $-$ 1.7, $-$ 2.0)\\
		Nitrogen prescription & No secondary N, Dopita13 prescription, Nicholls17 prescription, Groves04 prescription\\
		\hline
	\end{tabular}
\end{table*}

In order to investigate the influence of each input parameter on the final model grids, we vary the following parameters in our models:
\begin{enumerate}
\item Density of hydrogen. Typical NLRs have n$\rm _H$ $\gtrsim$ 10$^2$ $-$ 10$^3$ cm$^{-3}$ (\citealp{2008ARA&A..46..475H}), and in most cases we can assume that for the line-emitting regions inside the clouds, the electron density is approximately the same as the hydrogen density: n$\rm _e$ $\approx$ n$\rm _H$. Still, to account for all possible scenarios, we set four density values for comparison: log(n$\rm _H$/cm$^{-3}$) = 1, 2, 3, 4.
\item Input spectral energy distribution (SED). A typical Seyfert has an SED consisting of a radio component due to synchrotron radiation, an IR excess caused by re-emission of dust grains, a UV bump representing the disk component, an soft X-ray excess and finally a power-law component most noticeable in hard X-ray (\citealp{1978Natur.272..706S, 1982ApJ...254...22M, 2007ASPC..373..121D, 2008ARA&A..46..475H, 2009MNRAS.394..443M}). For practical usage, most often a featureless power-law component combined with cut-offs in low and high energy ends is good enough to derive sensible results. Here, we consider both SEDs of a composite type and those of a power-law type. The composite SEDs we use are based on the model described by \cite{2006hbic.book.....F} and \cite{2014MNRAS.437.2376R}. And the power-law SEDs we adopt have two different set-ups: one is from the built-in command of CLOUDY (\citealp{2006hbic.book.....F}), and the other is similar to the one adopted by \cite{2004ApJS..153....9G}. We postpone the detailed description of these SEDs to Section 4.2.
\item Production of secondary elements. BPT diagnostic is based on emission line ratios and thus is sensitive to extra production of elements. Net yield of nitrogen from the CNO cycle in pre-enriched stars could have a non-negligible impact on the [N~II]-based BPT diagnostics. Besides, the cooling effects from extra nitrogen and carbon might also change the loci of the upper boundary of AGN regions on all BPT diagrams. By the same token, the primary production of helium could also contribute to the cooling and thus shape the boundary. The formation of the latter is relatively certain than the former, as the actual contributors and procedure of the secondary elements are still under debate (\citealp{2000ApJ...541..660H}). Nevertheless, we have compared different methods to take the secondary elements into account. They all show a similar behavior but to different degrees. We put the detailed discussion in the corresponding subsection.
\end{enumerate}

In Table.~\ref{tab:model_sum} we give a summary of the models we use. In general, as we will see, the photoinoization model grids tend to fold themselves and produce degeneracy at high metallicity, regardless of the ionization parameter. This behavior resembles the well studied model grids of the H~II region, which explains theoretically the existence of its upper boundary (Ke01 and Ka03 lines). So we would expect the similar physics is behind this phenomenon, despite happening in a new parameter regime. Qualitatively, the fact that this boundary is higher than the H~II boundary can be explained by the same reason why AGN regions are higher in BPT diagrams than H~II regions: forbidden lines like [N~II]$\lambda \lambda$6548, 6583 and [S~II]$\lambda \lambda$6716, 6731 are created by the collisionally excited N$^+$ and S$^+$. These ions are mostly found in the partially ionized region of hydrogen in the cloud, where they can coexist with hot electrons in large number. The radial extent of this partially ionized region hinges on the hardness of the ionizing spectrum. Therefore, a harder SED like one of AGN would presumably produce a higher ratio of [N~II]$\lambda$6583/H$\alpha$ and [S~II]$\lambda \lambda$6716, 6731/H$\alpha$ than that produced by a softer SED like one of OB stars.

If we can make a quantitative connection between the position of the upper boundary in BPT diagrams with certain set of physical parameters, we may, by analogy to what people have done to H~II regions, go back to constrain the input parameters for observational data. This is our main goal of this paper.

\subsection{Impact of hydrogen density}

\begin{figure*}
	\includegraphics[width=0.9\textwidth]{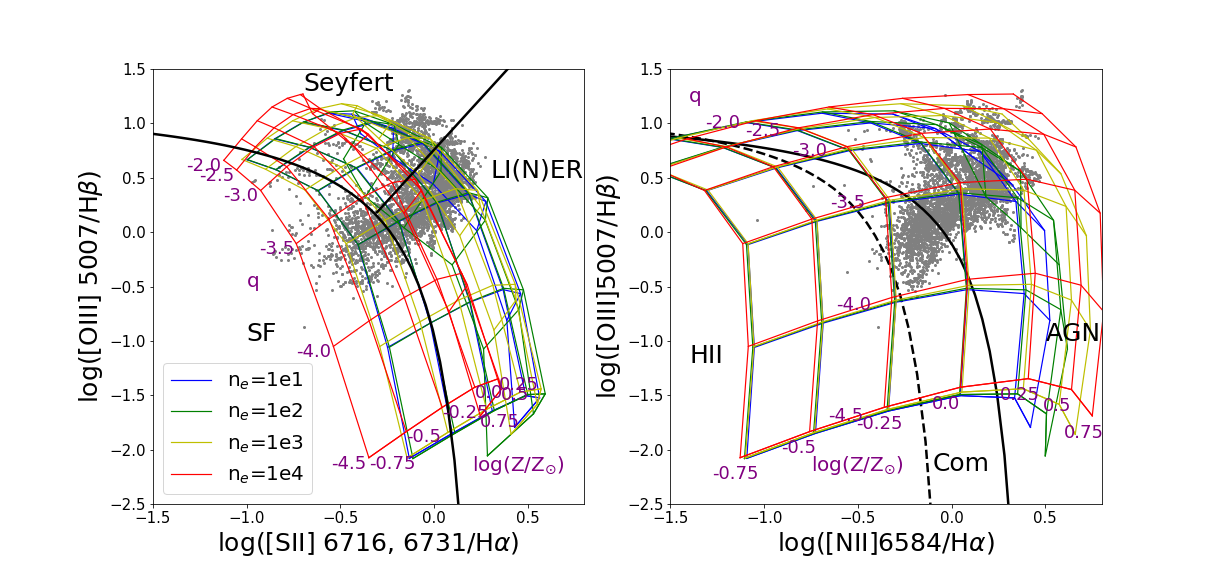}
	\caption{Photoionization grids generated by {\tt CLOUDY} with different hydrogen densities in the [S~II] and [N~II] BPT diagrams. Densities used for the models range from 10 cm$^{-3}$ to 10$^4$ cm$^{-3}$. The input SED for these grids is a composite AGN SED with an accretion disk temperature of 4.2 $\times$ 10$^5$ K, an optical to X-ray ratio $\alpha_{\rm ox}$ = $- 1.35$, a UV slope of $-0.3$, and an X-ray slope of $-1.0$. For each constant metallicity (or ionization parameter) line of the model grid with the highest density (in red), the value of the parameter is shown beside the line. Part of the model with the lowest density (in blue) is not plotted, which is subject to very low temperature and produces line ratios unrealistically low and sensitive to the stopping criteria of {\tt CLOUDY}. This happens at the highest metallicity, log(Z/Z$_{\odot}$) = 0.75.}
    \label{fig:hden}
\end{figure*}

Fig.~\ref{fig:hden} shows the photoionization grids with different hydrogen densities in both [S~II] and [N~II] BPT diagrams. Interestingly, almost all model grids seem to fold themselves at very high metallicity, starting around Z = 1.78 Z$_{\odot}$ (Z$_{\odot}$ = (O/H)$_{\odot}$ = 10$^{-3.31}$, cf.~\citealp{2010Ap&SS.328..179G}), creating an upper right boundary on both BPT diagrams. Without further tuning of input parameters, the upper boundaries of model grids already lie very close to that shown by the MaNGA data. Still this boundary has dependence upon the hydrogen density to some degree.

In the low density regime ($1 {\rm cm}^{-3} < n_{\rm H} < 1000 {\rm cm}^{-3}$), the positions of the grids are very similar to each other in the [S~II] BPT diagram, which only moves upwards slightly as the density increases. In the [N~II] BPT diagram, however, the effect of increasing density is more obvious. As density becomes even higher and reach 10$^4$cm$^{-3}$, the grids on the [S~II] BPT diagram start to move leftwards: while the [O~III]$\lambda$ 5007/H$\beta$ keeps rising, the [S~II]$\lambda \lambda$ 6716, 6731/H$\alpha$ has a conspicuous decrease. Meanwhile, the grids on the [N~II] BPT diagram keep on going upper-rightwards. The behavior of the model grids can be qualitatively explained by the different critical densities of forbidden lines, above which the line strength would drop significantly due to the effect of collisional de-excitation. The [S~II]$\lambda \lambda$ 6716, 6731 have the smallest critical density of 2 $\times$ 10$^3$ cm$^{-3}$, while the [N~II]$\lambda$ 6583 and [O~III]$\lambda$ 5007 have critical densities of 6.6 $\times$ 10$^4$ and 7 $\times$ 10$^5$ respectively. Within the range of density we consider, only the [S~II] doublet will surpass its critical density and show obvious drop in line intensities.

\begin{figure}
	\includegraphics[width=0.5\textwidth]{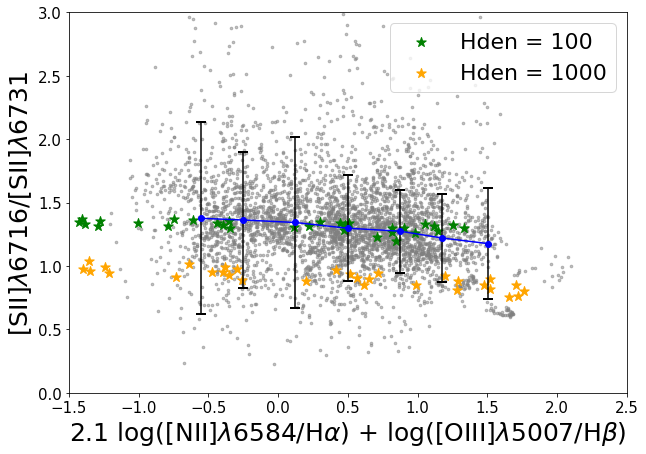}
	\caption{The ratios of [S II] doublet of our sample spaxels (grey points) and two {\tt CLOUDY} models with different densities (100 cm$^{-3}$ for green star symbols and 1000 cm$^{-3}$ for orange star symbols), as a function of the analog distance in the upper right direction in the [N II] BPT diagram.}
    \label{fig:s2r}
\end{figure}

Generally speaking the NLRs can have a broad distribution of electron densities from 1 cm$^{-3}$ to 10$^4$ cm$^{-3}$. But there are few cases where the density is below 1 cm$^{-3}$ or exceeds 10$^3$ cm$^{-3}$, which means in the ordinary density regime the photoionization model grids only have a weak dependence on the density. To further verify this point, we estimate the electron density of our sample galaxy spaxels using the [S~II]$\lambda$ 6716 / [S~II]$\lambda$ 6731 ratio, assuming a constant electron temperature of 10$^4$ K. Our result shows that the median value of [S~II]$\lambda$ 6716 / [S~II]$\lambda$ 6731 is 1.31 for our sample. The 16 percentile and 86 percentile values are 1.05 and 1.65, respectively. Using the {\tt nebular.temden} routine (\citealp{1987JRASC..81..195D, 1995PASP..107..896S}) in {\tt PyRAF}, we roughly estimate the electron densities of our sample spaxels (within 1 $\sigma$) to range from $< 10$ cm$^{-3}$ to 4.9 $\times$ 10$^2$ cm$^{-3}$, with a median value of 110 cm$^{-3}$. This range is close to what we have expected of NLRs despite some systematic errors that might come along with temperature variation. In Fig.~\ref{fig:s2r} we plot the [S~II] doublet ratios versus the analog distance measured in the upper right direction in the [N~II] BPT diagram for our sample and photoionization models. The [S~II] doublet ratio shows a slightly decreasing trend towards the upper right direction in the [N~II] BPT diagram, which implies a small increase in the electron density. If we look at the ratios of [S~II] doublet directly from our {\tt CLOUDY} models, however, the model with a density of 100 cm$^{-3}$ is in best agreement with the median of our data. Therefore, in the following calculation we will always assume a density of 100 cm$^{-3}$ to best represent our data.

\subsection{Impact of varying AGN SED}

A typical AGN SED can be decomposed into five parts with increasing energy: a possible radio component from synchrotron emission in the jet, an IR component produced by dusts, a big blue bump from the accretion disk, a soft X-ray excess, and a power-law component which is most noticeable at hard X-ray \citep{1978Natur.272..706S, 1982ApJ...254...22M, 2007ASPC..373..121D, 2008ARA&A..46..475H, 2009MNRAS.394..443M}. It is often convenient to describe the whole continuum with an analytical function:
\begin{equation}
    \text{F}_\nu = \nu ^{\alpha _{\rm uv}}\text{exp}(-\text{h}\nu /\text{kT}\rm _{BB})\text{exp}(-\text{kT}\rm _{IR}/\text{h}\nu)+a\nu ^{\alpha_{\rm x}} ,
    \label{eq_sed}
\end{equation}
where $\alpha \rm _{uv}$ determines the UV slope and $\alpha \rm _x$ the X-ray slope. T$\rm _{BB}$ sets the location of the big blue bump while T$\rm _{IR}$ sets an infrared cutoff for it. The coefficient $a$ determines the optical-to-X-ray slope $\alpha \rm _{ox}$, which is not explicitly shown in this equation. At low energy the last term would be omitted while at high energy F$_\nu$ would be assumed to be proportional to $\nu ^{-2}$. 

\begin{figure}
	\includegraphics[width=0.5\textwidth]{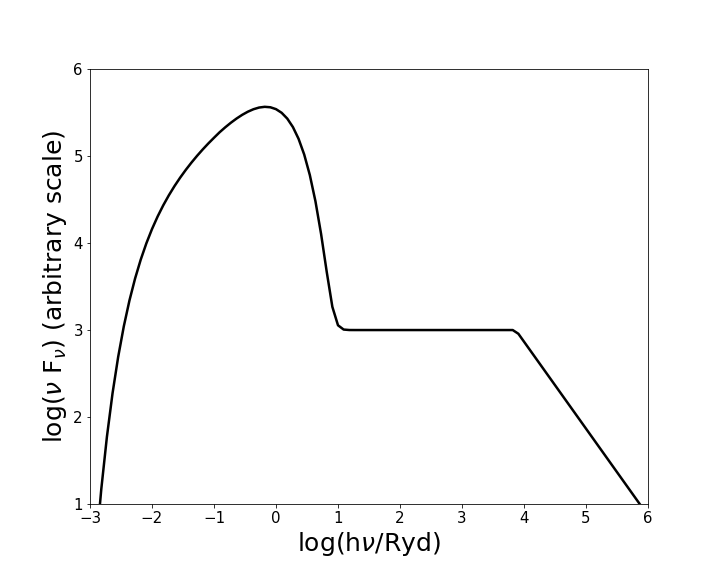}
	\caption{The shape of the composite AGN SED we use as input for {\tt CLOUDY}. The vertical axis has not been scaled.}
    \label{fig:SED_com}
\end{figure}

In Fig.~\ref{fig:hden} we use an AGN SED of this type, similar to the optimal SED that \cite{2014MNRAS.437.2376R} adopted, with $\alpha_{\rm uv}$ = $-0.3$, $\alpha \rm _{x}$ = $-1.0$, $\alpha_{\rm ox}$ = $-1.35$, T$\rm _{BB}$ = 4.2$\times$10$^5$ K and T$\rm _{IR}$ = 1.58$\times$10$^3$ K. Fig.~\ref{fig:SED_com} shows the shape of such a continuum. To vary the AGN SED, one can change the weight of different parts, which will result in a large number of combinations.

Observationally, people have found that the big blue bump does not necessary exist especially for LI(N)ERs (\citealp{2008ARA&A..46..475H}). The optical-to-UV slope $\alpha_{\rm uv}$ can vary from $-1$ to $-2.5$ for low luminosity AGNs and becomes around $-0.5$ to  $-0.7$ for luminous AGNs (\citealp{2001AJ....122..549V, 2005ApJ...619...41S}). Also the optical-to-X-ray slope $\alpha_{\rm ox}$ might vary among AGNs, with high luminosity AGN like quasars and Seyferts having $\alpha_{\rm ox}$ $\approx$ $-1.4$ $\sim$ $-1.2$ and low luminosity AGN like LI(N)ERs having $\alpha_{\rm ox} < -1.0$ (\citealp{1993nag..conf.....B, 1989ApJ...339..674M}).

For practical usage, one usually wants to isolate the part in the AGN SED that is most relevant to the problem. In most cases a single power-law component can well represent the influence of energy input from AGN since it characterizes the ionizing energy regime ( $> 1$ Ryd) in the spectrum. Certainly one would also need to specify the low and high energy cutoffs while using a single power-law SED in order to avoid unphysical conditions.

\begin{figure}
	\includegraphics[width=0.5\textwidth]{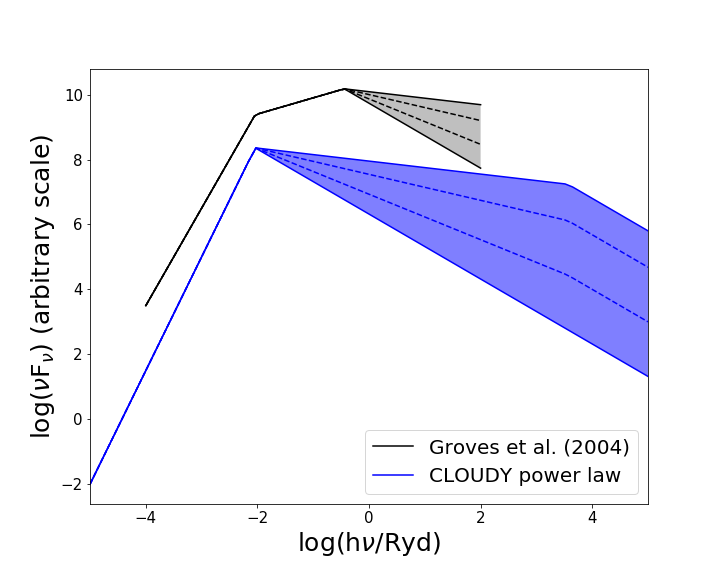}
	\caption{Input AGN SEDs for photoionization models that are used by \protect\cite{2004ApJS..153...75G} and {\tt CLOUDY} code, respectively. The intensity (vertical axis) of each set of SEDs is intentionally shifted for convenience of comparison. Each SED is divided into three segments. There are four different power-law slopes (or $\alpha$) at the high energy end for Groves's SED: $-$ 1.2, $-$ 1.4, $-$ 1.7, and $-$ 2.0. For the SED used by {\tt CLOUDY}'s {\tt table powerlaw} command, we vary the power-law slope in the intermediate energy range.}
    \label{fig:SED_sum}
\end{figure}

\begin{table}
	\centering
	\caption{Model SED summary}
	\label{tab:sed_sum}
	\begin{tabular}{lccr}
	    \hline
	    \hline
	    \multicolumn{2}{c}{Groves04} & \multicolumn{2}{c}{\tt CLOUDY}\\
		\hline
		h$\nu$ (Ryd) & $d\log F_\nu / d\log \nu$ &  h$\nu$ (Ryd) & $d\log F_\nu / d\log \nu$\\
		\hline
		1.00 $\times$ 10$^{-4}$ & + 2.00 & 1.00 $\times$ 10$^{-8}$ & + 2.50\\
		9.115 $\times$ 10$^{-3}$ & $-$ 0.50 & 9.115 $\times$ 10$^{-3}$ & $-$ 2.0 $\sim$ $-$ 1.2\\
		0.36 & $-$ 2.0 $\sim$ $-$ 1.2 & 3676. & $-$ 2.00\\
		91.35 & $-$ & 7.40 $\times$ 10$^{6}$ & $-$\\
		\hline
	\end{tabular}
\end{table}

People have used different energy cutoffs for their power-law SEDs for different reasons. Here we apply both the one that is adopted in \cite{2004ApJS..153...75G} and the one that is built in the {\tt CLOUDY} code (\citealp{2006hbic.book.....F}). The setting of the input SED is summarized in Fig.~\ref{fig:SED_sum} and Table.~\ref{tab:sed_sum}. Basically the {\tt CLOUDY} type SEDs have a wider energy range. Whereas both sets of SEDs have accounted for the self-absorbed synchrotron radiation at the low energy end by using a steep slope, {\tt CLOUDY} type SEDs go much further into the high energy domain and use a fixed power-law slope for the high energy end.

\begin{figure*}
	\includegraphics[width=0.9\textwidth]{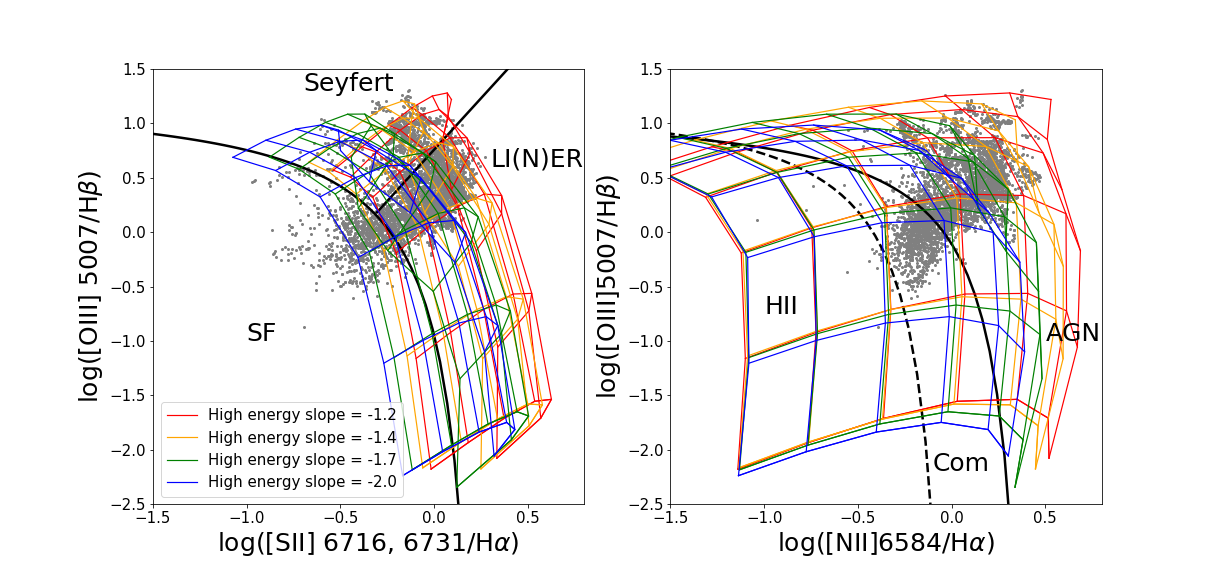}
	\includegraphics[width=0.9\textwidth]{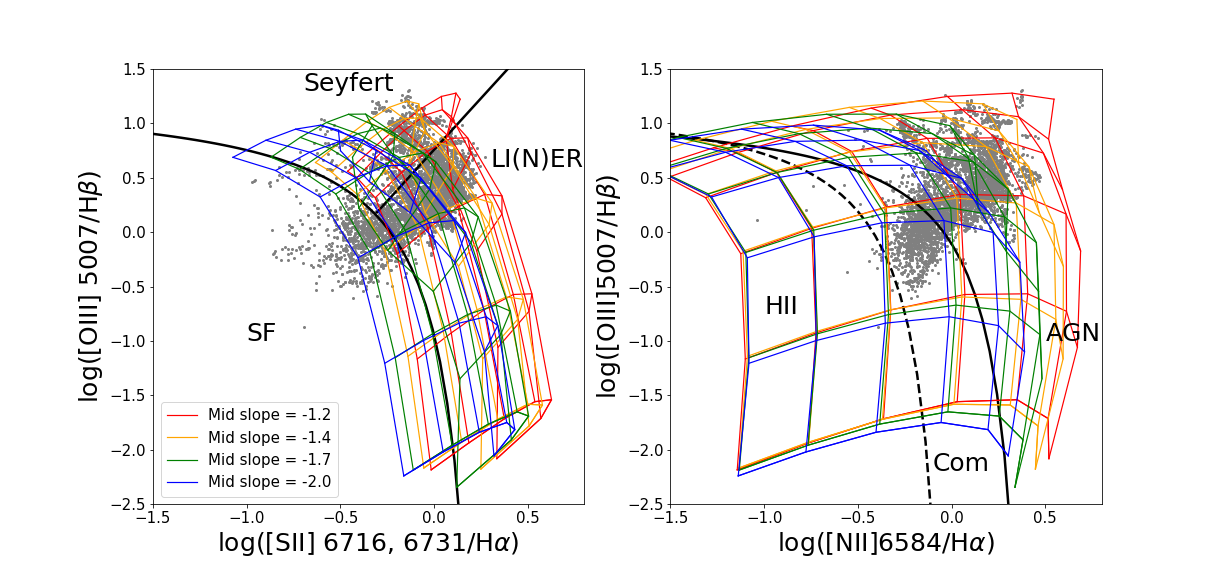}
	\caption{Photoionization grids generated by different input AGN power-law SEDs. Upper panels: Groves' type SEDs. Lower panels: {\tt CLOUDY} type SEDs. The ranges of metallicity and ionization parameter are the same as in the previous plots. Part of the model with the softest SED (in blue) is not plotted, which is subject to very low temperature and produces line ratios unrealistically low and sensitive to the stopping criteria of {\tt CLOUDY}. This happens at the highest metallicity, log(Z/Z$_{\odot}$) = 0.75.}
    \label{fig:sed}
\end{figure*}

Despite the fact that the two sets of power-law SEDs use different low and high energy slopes and cut off at different energies, they yield very similar results. Fig.~\ref{fig:sed} show the grids generated by different input SEDs in both [S~II] and [N~II] BPT diagrams. As the power-law component becomes flatter, or equivalently as the SED becomes harder, boundaries of model grids are shifted towards the upper-right direction, implying increases in all four line-ratios at high metallicity. The behavior of output model grids with different SEDs can be explained as the change of the relative strength of collisional excited lines and recombination lines of hydrogen. The key feature, the hardness of the spectra, is roughly determined by the shape of the input SED between 1 Ryd and 100 Ryd. This explains why two SEDs with very different forms can produce similar output grids. By comparing Fig.~\ref{fig:hden} with Fig.~\ref{fig:sed}, one can see that a power-law SED yields a nearly identical model grid to that produced by a much more complicated SED. The model grids with composite SED in Fig.~\ref{fig:hden} can be well approximated by a power-law SED with a power-law index of $\sim$ $-$ 1.5. A small but discernible difference, however, does occur at very high metallicity and ionization parameter, where the power-law SEDs with flat power-law indices seem to be distorted. This actually is the feature of degeneracy at high ionization parameter, as we will see later. Model grids from a power-law SED and a composite SED we use here would begin to deviate from each other significantly at very high q value, although both would have degeneracy.

\begin{figure}
	\includegraphics[width=0.5\textwidth]{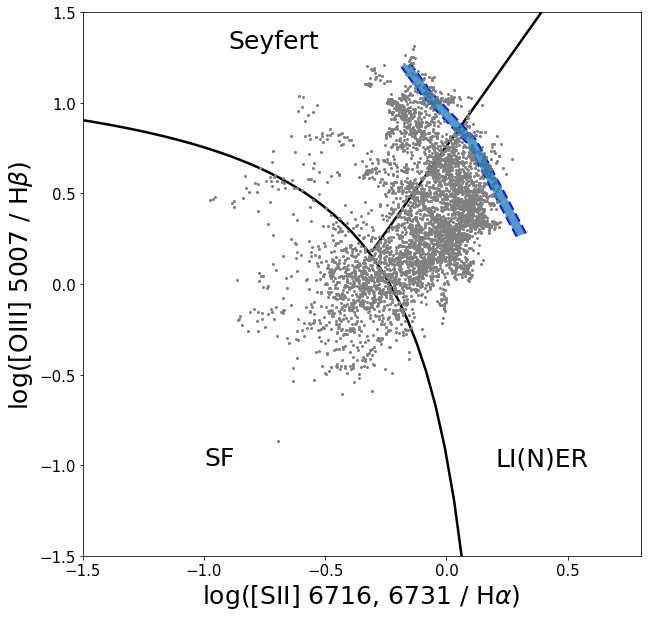}
	\caption{Allowed positions for the upper-right boundary in the [S~II] BPT diagram given the measurement uncertainties of emission lines, as is shown in the shaded blue. This region corresponds to a variation in the power-law index close to 0.05 for the input SED.}
    \label{fig:bound_err}
\end{figure}

Judging from the [S~II] BPT diagram, an input Groves type SED with a high energy slope of $-$ 1.4 matches the data best. This is consistent with the case of a {\tt CLOUDY} type SED of which the middle energy slope is $-$ 1.4. While for the [N~II] BPT diagram, it is hard to constrain the shape of SED as there is no well defined boundary. Since we have shown that the hydrogen density has barely any effect on the position of the upper-right boundary, we can constrain the power-law index of a typical AGN to be $\sim$ $-$ 1.4 using the distribution of MaNGA spaxels in the [S~II] BPT diagram. The measurement errors of the four emission lines near the observed boundary introduce a 1$\sigma$ uncertainty of 0.020 dex in determining the position of this boundary (A correction factor of 1.25 has been applied using the MaNGA repeat observations; cf.~\citealp{2019AJ....158..160B}). The corresponding uncertainty in the power-law slope of the input SED is approximately 0.05. In Fig.~\ref{fig:bound_err} we plot the allowed region for the boundary. One caveat is that the intrinsic variation of the AGN SED (if larger than the range we derive) might blur the distribution of the data points in the [S~II] BPT diagram. We can only constrain the hardest SED in our sample with this upper-right boundary. It is possible that there are some softer SEDs having lower boundaries that are hidden inside the AGN sequence we see in BPT diagrams. But since there is no obvious piling up of the data points except at the location of the upper-right boundary, we think the impact of this intrinsic variation should be small.

\subsection{Impact of secondary elements}

Nitrogen, as a secondary element, does not simply scale with the oxygen abundance like many others. And since it is used in the [N~II] BPT diagnostics, it is likely that the secondary nitrogen can have non-negligible effects on the model grids. How to calculate the amount of secondary nitrogen, however, remains debated. Nitrogen is produced by the CNO cycle inside massive and intermediate-mass stars. While the secondary nitrogen mainly comes from the initial oxygen and carbon incorporated into a star at its formation, the primary nitrogen is created through oxygen and carbon newly formed inside the star, and thus is independent of the initial chemical abundance. Therefore, one would expect the contribution from secondary nitrogen becomes more and more important as metallicty increases. Ideally the final N/O would be proportional to C/H as well as O/H for a secondary origin of nitrogen. However the stellar evolution and the release of the nitrogen into the ISM could have dependence on metallicity, thus complicating the situation (\citealp{2000ApJ...541..660H}).

Here we compare three different ways to account for the secondary elements. The first one is our default prescription taken from \cite{2004ApJS..153....9G} (hereafter Groves04). Eq.~\ref{eq_gro} describes how the nitrogen to oxygen ratio changes with metallicity according to this prescription.

The second nitrogen prescription we compare here comes from \cite{2017MNRAS.466.4403N} (hereafter Nicholls17). They use an equation with a similar form to express the relation between nitrogen abundance and metallicity:
\begin{equation}
    \rm N/O = 10^{-1.732} + 10^{2.19 + log(O/H)} .
    \label{eq_nic}
\end{equation}
One can see that this prescription provides both lower primary nitrogen abundance (indicated by the first term), and lower secondary nitrogen abundance (indicated by the second term) at fixed metallicity.

\begin{figure}
	\includegraphics[width=0.5\textwidth]{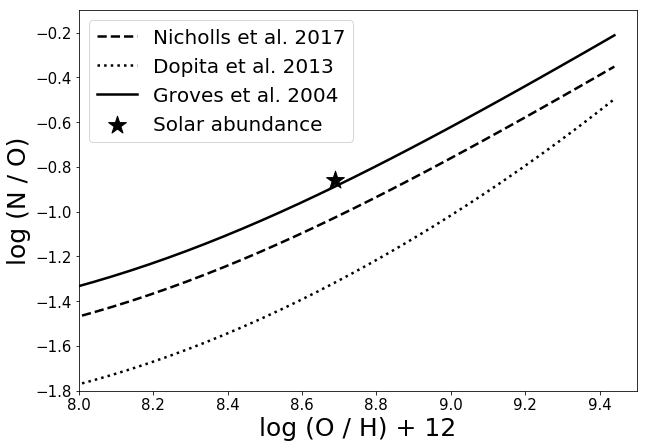}
	\caption{Comparison of three different nitrogen prescriptions used in this paper. The solar abundance is indicated by the star symbol.}
    \label{fig:noc}
\end{figure}

The third nitrogen prescription we choose is from \cite{2013ApJS..208...10D} (hereafter Dopita13). They obtain the relation with a set of observational data from \cite{1998AJ....116.2805V}. Here we refit the relation using the data points provided by \cite{2013ApJS..208...10D} with a quadratic function:
\begin{equation}
    \text{N/O} = 0.0096 + 72\cdot \text{O/H} + 1.46\times 10^4 \cdot \text{(O/H)}^2 .
    \label{eq_dop}
\end{equation}
This method gives the smallest N/O at all metallicities among the three methods.

A comparison of these three nitrogen prescriptions is shown in Fig.~\ref{fig:noc}. The effect of metal depletion has been tuned to be the same for all three methods. Thus any discrepancy in the output should directly result from the initial abundance setting. Since for now we are chiefly concerned with the upper-right part of the models in BPT diagrams, where metallicity is sufficiently high, we expect the resulting differences are mainly determined by the parts of these curves with log(O/H) + 12 $\gtrsim$ 9.0.

For another important secondary element, carbon, we set its abundance to be always 0.6 dex larger than the nitrogen abundance for the first two sets of models, following the observational data in \cite{2004AJ....128.2772G}, resulting in a constant C/N. For the third set of models, however, we require log(C/N) = 1.03 dex, in order to be consistent with \cite{2013ApJS..208...10D}.

\begin{figure*}
	\includegraphics[width=0.9\textwidth]{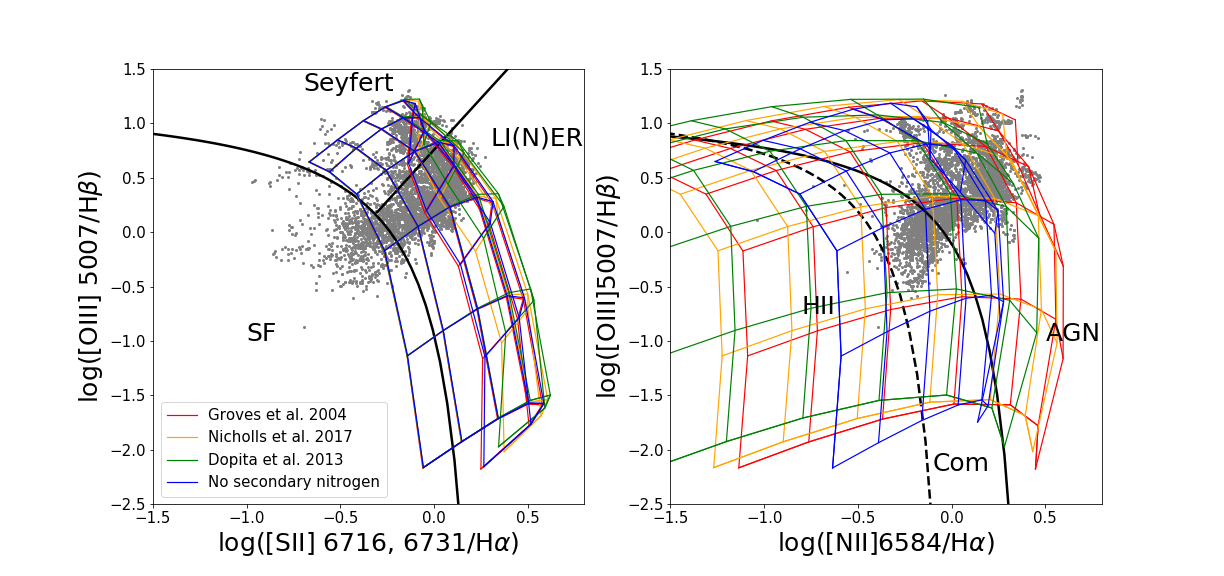}
	\caption{Photoionization models with and without secondary elements. The hydrogen density is set to be 10$^2$ cm$^{-3}$. And the input SED is a power-law SED with a power-law index of $-$ 1.4 (A Groves' type).}
    \label{fig:second_n}
\end{figure*}

\begin{figure*}
	\includegraphics[width=0.9\textwidth]{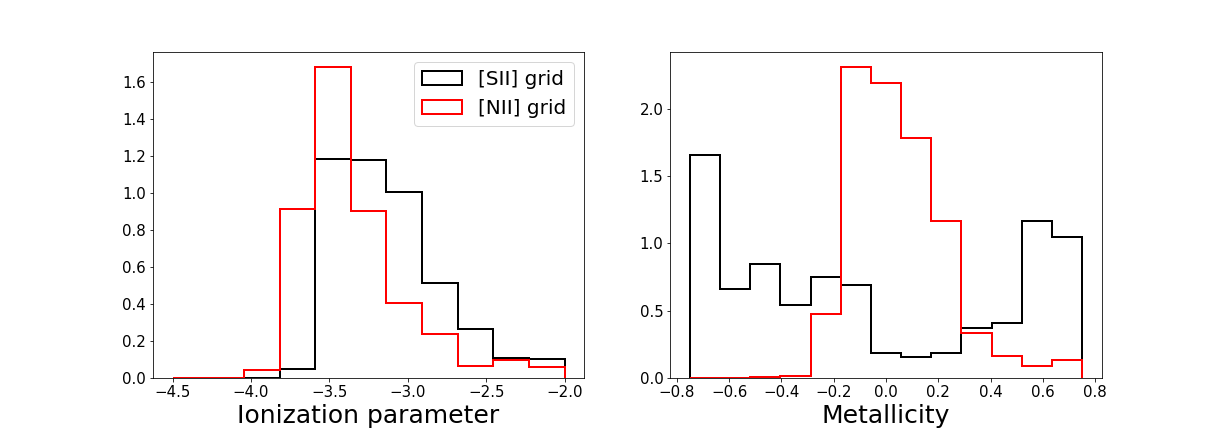}
	\caption{Predictions of ionization parameter (left panel) and metallicity (right panel) using a Groves04 type model grid in [S~II] and [N~II] BPT diagram respetively. The histograms show the normalized density distribution of our sample spaxels. The model grid is interpolated so that the ionization parameter q ranges from $-$ 4.5 to $-$ 2.0 with a step size of 0.25 dex, and the metallicity log (Z / Z$_\odot$) ranges from $-$ 0.75 to 0.75 with a step size of 0.125 dex. For each spaxel and for each diagram, we find the nearest node of the model on the BPT plane and use the corresponding values as the predictions.}
    \label{fig:predict}
\end{figure*}

Besides these three prescriptions, people have come up with many other relations with different data sets \citep[e.g.][]{1993MNRAS.265..199V, 1998AJ....115..909S, 2020ApJ...890L...3S}. Among these relations, the Dopita13 and Groves04 relations appear as lower and upper bounds, respectively. While the Nicholls17 relation appears as a median one. We note that the exact relations could depend on the properties of galaxies or regions. Nevertheless, these three prescriptions already cover a large range in the allowed region of the parameter space spanned by N/O and O/H ratios.

Note that we use these three methods as the input relations, but the output relations could be different, as our photoionization models include metal depletion process. So what Fig.~\ref{fig:second_n} shows are the pre-depletion prescriptions. In our models oxygen depletes, but not nitrogen. We expect the post-depletion N/O vs. metallicity curves to be systematically higher than the pre-depletion ones.

In Fig.~\ref{fig:second_n}, we plot model grids with and without secondary elements. In the [S~II] BPT diagram, it can be seen that the influence of secondary elements is modest. Near the upper-right boundary, models with secondary elements give slightly smaller values of [S~II]$\lambda \lambda$ 6716, 6731 / H$\alpha$, while the differences in [O~III]$\lambda$ 5007 / H$\beta$ are even smaller. Extra cooling brought by secondary elements at high metallicity might lead to this small effect. To verify this point, we check the electron temperature near the upper-right boundary, where Z $\approx$ 1.78 Z$\odot$. It turns out that including secondary elements do lower the overall electron temperature, but the relative change in temperature is quite small, ranging from 1 \% to 2 \% from high ionization parameter to low ionization parameter.

In the [N~II] BPT diagram, the effect is much more conspicuous. Being a secondary element itself, the extra nitrogen has shifted the boundary greatly towards the right. With the secondary nitrogen, the degeneracy at very high metallicity Z $\approx$ 1.78 Z$\odot$ is removed, as the model grids are stretched horizontally. We can not identify a clear sign of an upper-right boundary in this case. This explains why the distribution of data points in the [N~II] BPT diagram seems less compact than that in the [S~II] BPT diagram, for the spaxels are only loosely bounded by a very high boundary on the plane. Models with larger overall nitrogen abundance are able to enclose more data points distributed on the upper-right position. Among all three models, the one with the Groves04 nitrogen prescription can explain the most data in our sample, making it the best-fitting model. Still it is worth noting that the difference between this model and the model using the Nicholls17 relation is very small. A possible explanation for the differences between the three models is that the Groves04 relation is obtained with a sample including starburst nuclei, which might be more suitable to describe our data drawn from the center of galaxies. \cite{2017MNRAS.469..151B} suggests a positive correlation between the overall scale of the N/O vs. O/H relation and the total stellar mass of a galaxy. Similar effects might take place for regions inside a single galaxy with larger stellar mass densities, like the nuclear region, if we consider that the star formation history could vary inside a galaxy.

The non-folding feature of the model grids on the [N~II] BPT diagram could be useful for the determination of metallicity. However, the behavior of nitrogen at the median to low metallicity end significantly influences the calibration. As we can see in the right panel of Fig.~\ref{fig:second_n}, the Dopita13 method would predict systematically larger metallicity compared with the other methods. Even larger discrepancy would occur if one compares the predictions from the [S~II] BPT diagram and the [N~II] BPT diagram. Fig.~\ref{fig:predict} compares the predictions of ionization parameter and metallicity using a Groves04 type model grid. The predictions of ionization parameter shows much better consistency compared with those of metallicity. By eye it is obvious that in the [S~II] BPT diagram, even the lowest metallicity cannot bound all the spaxels. One possible explanation is that the lower spaxels in our sample have more contamination from non-AGN sources, and the impact is more significant in the [S~II] BPT diagram. Another factor that could contribute to the discrepancy is the variation of N/O among galaxies. \cite{2017MNRAS.469..151B} and \cite{2020ApJ...890L...3S} have shown that the relationship between N/O and metallicity has dependency on global properties of galaxies like stellar masses and star formation efficiencies. If the lower spaxels on the [N~II] BPT diagram have higher N/O ratios than those prescribed by these relations at the low metallicity end, the tension between the two diagrams could be partly solved. Still this argument alone cannot fully settle the issue and further analyses with the sample spaxels are needed. Later in this section we will combine the [S~II] and [N~II] BPT diagrams to show that the first scenario is more likely to be the case.

\subsection{Behavior at high ionization parameter}

\begin{figure*}
	\centerline{\includegraphics[width=1.1\textwidth]{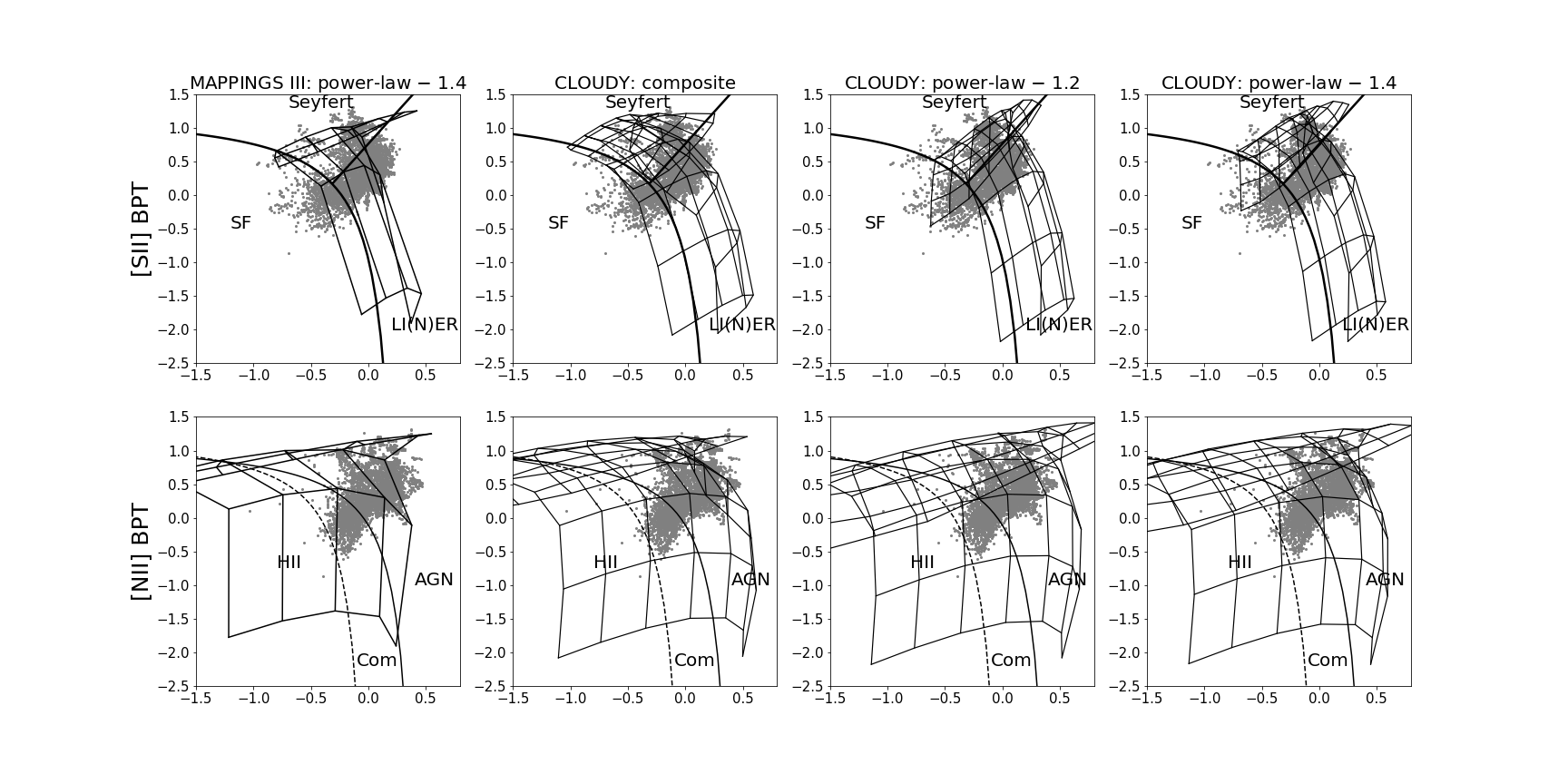}}
	\caption{Photoionzation model grids reaching high ionization parameters. Four sets of model grids are displayed, from left to right, they are: (1) Model grids generated by {\tt MAPPINGS III} with a power-law SED of which the index is $-$ 1.4; (2) Model grids generated by {\tt CLOUDY} with a composite AGN SED of which the parameter set is identical to what we use previously; (3) {\tt CLOUDY} model grids with a power-law SED of which the power-law index is $-$ 1.2; (4) {\tt CLOUDY} model grids with a power-law SED of which the power-law index is $-$ 1.4. The form of power-law SED used here is Groves' type. For {\tt CLOUDY} models, q ranges from $-$ 4.5 to 0.0, while Z ranges from 0.178 Z$_\odot$ to 5.62 Z$_\odot$ (-0.75 to 0.75 for log  (Z / Z$_\odot$)). For {\tt MAPPINGS III} models, q ranges from $-$ 4.0 to 0.0, and Z ranges from 0.25 Z$_\odot$ to 4.0 Z$_\odot$.}
    \label{fig:selong}
\end{figure*}

Using the code, {\tt MAPPINGS III}, \cite{2004ApJS..153...75G} noticed that for a dusty and isobaric cloud, its emission line ratios would display degeneracy when the ionization parameter exceeds $-$ 2.0. In Fig.~\ref{fig:selong} we show their results as well as what we obtain from CLOUDY (with different input SEDs) when the high ionization parameter regime is included.

It is clear that all models shown in Fig.~\ref{fig:selong} have upper-left boundaries as a result of degeneracy in line ratios at the high q regime. Besides, it is interesting that the position of this boundary is also sensitive to the shape of the photoionization source. However, unlike the case of the upper-right boundary, a composite SED would produce a remarkably different upper-left boundary from that generated by a power-law SED with an index of $-$ 1.4, making this regime a potential complement to better discriminate different SEDs. One remaining question is that whether this regime is adequately populated with the current data we have. Our sample does not fully populate the Seyfert region as we enforce large D$\rm _n$(4000) during the sample selection. And with the current BPT diagrams it is hard to tell if a continuous distribution exists in the upper-left direction.

\begin{figure*}
	\centerline{\includegraphics[width=1.1\textwidth]{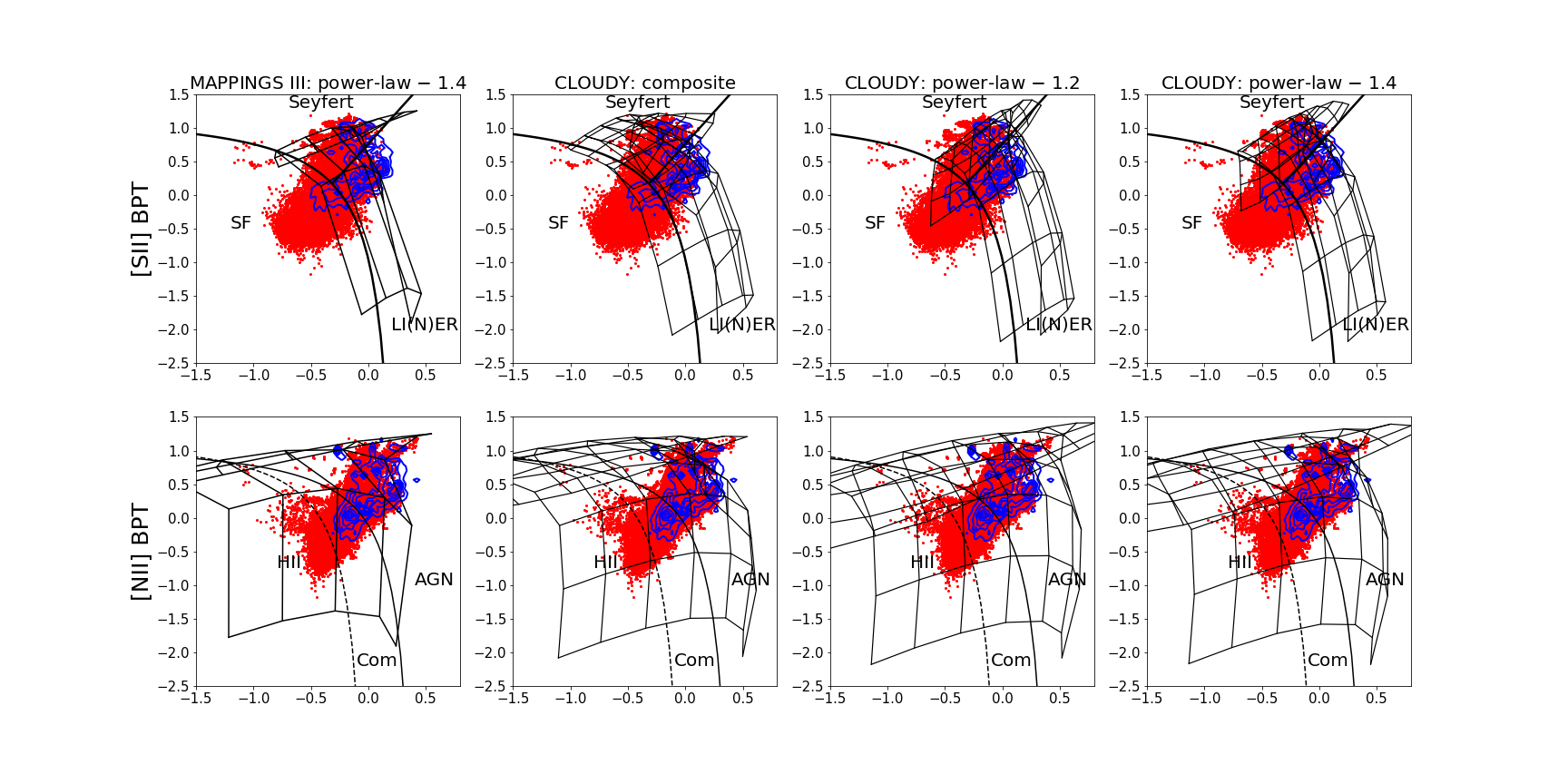}}
	\caption{Same as Fig.~\ref{fig:selong}, despite that spaxles with $1.5 < D_{\rm n}(4000) \leq 1.8$ are included, which are shown in red. While the spaxels with $D{\rm _n}(4000) > 1.8$ are plotted as blue contours.}
    \label{fig:selonglow}
\end{figure*}

In Fig.~\ref{fig:selonglow} we add spaxels with $1.5 < D_{\rm n}(4000) \leq 1.8$ into the two kinds of BPT diagrams. Solely judging from the [S~II] BPT diagram, the {\tt CLOUDY} model with a power-law index of $-1.4$ fits the data best, as the spaxels are well confined by the upper-left boundary. The {\tt CLOUDY} model with a composite SED also bounds the data points well, despite that the upper left boundary is a little more leftwards. Turning to the [N~II] BPT diagrams, while both power-law models can still enclose all data points, the other two models miss a number of points at high metallicity and ionization parameter. This time in the case of a {\tt CLOUDY} model with a power-law SED, the data points do not hit the upper left boundary. Thus, it is less likely that there is a continuous distribution towards the regime with very high ionization parameter.

\begin{figure*}
	\includegraphics[width=0.95\textwidth]{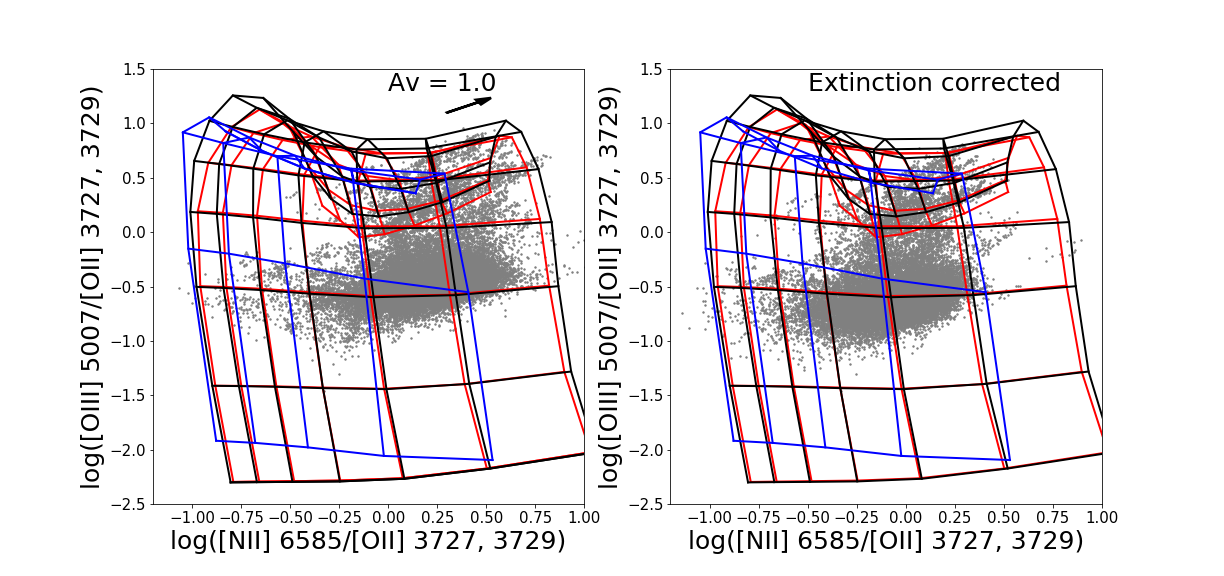}
	\caption{[O~III] $\lambda$ 5007/[O~II] $\lambda \lambda$ 3727, 3729 versus [N~II] $\lambda 6584$/[O~II] $\lambda \lambda$ 3727, 3729 diagram. Three different models from different codes and input SEDs are shown here: Red grid: {\tt CLOUDY} model with a power-law index of $-$ 1.2; Black grid: {\tt CLOUDY} model with a power-law index of $-$ 1.4; Blue grid: {\tt MAPPINGS III} model with a power-law index of $-$ 1.4. The background data points have D$\rm _n$(4000) > 1.5. On the left panel,  the data are not extinction-corrected, and the reddening vector is indicated by the arrow. On the right panel, we do the extinction correction based on Balmer decrements, assuming an intrinsic H$\alpha$/H$\beta$ ratio of 2.86.}
    \label{fig:o2bpt}
\end{figure*}

We investigate the [O~III]/[O~II] vs. [N~II]/[O~II] diagram in Fig.~\ref{fig:o2bpt}, of which the line ratios serve as good indicators of ionization parameter and metallicity. Since these combinations are sensitive to dust extinction, we use a reddening vector with A$\rm_V$ = 1 to indicate the direction of potential shift due to dust. Again we include all spaxels with D$\rm _n$(4000) > 1.5. In this diagram our photoionization models do not exhibit upper-right boundaries. Instead they produce upper boundaries, folding themselves at high [O~III]/[O~II] values. A small number of data points exceed the rightmost iso-metallicity lines with the highest metallicity. They are likely to be affected by the dust as their trails follow the reddening vector. In addition, the measurement errors in [O~II] doublets can also contribute to the shift. The bulk of the sample tend to cluster around q $\approx$ -3.5, which is consistent with our analysis with the [S~II] and [N~II] diagrams. Above this, we have small clusters of spaxels distributed closer to the upper boundary set by the two {\tt CLOUDY} models. However, there is an obvious gap between the data with the highest [O~III]/[O~II] ratio and the point where the grid wraps around, especially after the extinction correction. This means there are few spaxels that ever reach a high enough ionization parameter for the [O~III]/[O~II] ratio to wrap around. Otherwise, we would expect to see a continuous distribution up to an upper boundary and a pilling up of data points there. The fact we do not see this means the high q regime is not reached in a significant fraction of cases, at least in the MaNGA sample. Inspecting higher excitation lines could help the identification of high q spaxels, but this is beyond the scope of this work. Nevertheless it is of great significance if one can have a compilation of Seyfert data points confirmed having high q, which will not only place extra constraints on the input parameters such as the shape of SED, but also test the implicit assumptions made in different photoionization codes.

\subsection{Photoionization models in three-dimensional space}

\begin{figure}
    \animategraphics[autoplay,loop,width=0.5\textwidth]{3}{Section5/animation0/neww_rotation_}{0}{35}
    \caption{A 3D BPT diagram viewed at an elevation angle of 30 degrees. Our sample are shown as black points. An AGN model with a power-law index of $-$ 1.4 is displayed in grey. By comparison, a SF model with a input SED of a zero-age SSP with solar metallicity (from BC03, cf. \citealp{2003MNRAS.344.1000B}) is displayed in blue. The q of the SF model varies from $-$ 4.5 to $-$ 2.0, and the log(Z / Z$_{\odot}$) of the SF model varies from $-$ 2.3 to 0.75. The parameter space of the AGN model is the same as our default setting. We also interpolate the grids to smooth both surfaces. Click on this image in a PDF viewer to make it stop rotating.}
    \label{fig:3d_BPT}
\end{figure}

If we only consider model grids seen in the [N~II] and [S~II] BPT diagrams, ideally they should be projections of a continuous surface to two-dimensional space. One might ask when viewed in the three-dimensional space formed by log ([N~II] $\lambda$ 6584 / H$\alpha$), log ([S~II] $\lambda \lambda$ 6716, 6731 / H$\alpha$) and log ([O~III] $\lambda$ 5007 / H$\beta$), if the surface would provide extra constraint or show disagreement with the 2D results. 

In Fig.~\ref{fig:3d_BPT} we plot a rotating 3D BPT diagram with our sample and two different model surfaces. We notice that, although having some overlaps in 2D diagrams, the AGN surface and SF surface is clearly separated in the 3D line ratio space, implying the latter may be more powerful in identifying obscure cases. We can now view the original [N~II] and [S~II] BPT diagrams as two specific projections of this 3D one. For the [N~II] projection, the AGN surface is almost face-on, so there is no well-defined boundary in this diagram. While in the [S~II] projection, the AGN surface is nearly edge-on, causing the wrapping of the model to become very obvious in this diagram. This gives rise to the upper-right boundary we see.

With this diagram, we can check how well model surfaces match the data. The data points lie close to the AGN surface near the upper-right boundary we defined in the 2D BPT diagrams. Moving away from this surface, our sample forms a trail extending to the SF surface. This behavior provides support for our previous suspicion about the contamination of SF for the lower points in BPT diagrams. If that only the data points near the AGN surface are pure NLRs, then the true distribution of metallicity should be much narrower than what we see in Fig.~\ref{fig:predict}. Since the AGN surface is more parallel to the [O~III]/H$\beta$ axis, which is a proxy of the ionization parameter, the prediction for this parameter should be more reliable. The trail begins where our model predicts a low ionization parameter. This can be qualitatively explained by the fact that the influence of the central AGN should get weaker when the ionizing power decreases. If this scenario is true, then the lower tail of our sample does not really have low metallicity as indicated by the 2D AGN models. This picture is consistent with the idea that the NLRs are overall metal rich. It is worth noting that the transitioning trail is not a straight line. Its lower end does not directly reach the SF surface. Instead, it becomes gradually tangential to the high metallicity part of the SF surface. Note that we have applied a cut in D$\rm _n$(4000) for our sample, which might put a limit on the amount of SF we can have and thus cause this tangential behavior.

We see that combining the [S~II] and [N~II] BPT diagrams to perform a 3D analysis brings extra information that is hidden in the 2D cases. The requirement for the data to follow the surface can put tighter constraints on models. It is also possible that the degeneracy between different ionizing models in the 2D BPT diagrams is a result of projection effect and thus does not exist in the 3D line ratio space. Ideally if we consider more sets of line ratios simultaneously, more information is included and the diagnostics should become more accurate. However, as we have discussed before, the physical origins of different lines might differ dramatically. Thus one need to be cautious not to misinterpret the data when adding more lines. Nevertheless, it would be interesting to apply this analysis to other ionizing sources in future works.

\section{Shocks as an ionizing source}

\subsection{Shock models in the [S~II] BPT diagram}

Apart from photonionization, shocks are often considered as an important ionizing source especially for LI(N)ERs. We use a series of shock models from \cite{2008ApJS..178...20A} based on the {\tt MAPPINGS III} code and compare them with both our MaNGA sample and photoionization models.

\begin{figure*}
	\includegraphics[width=0.95\textwidth]{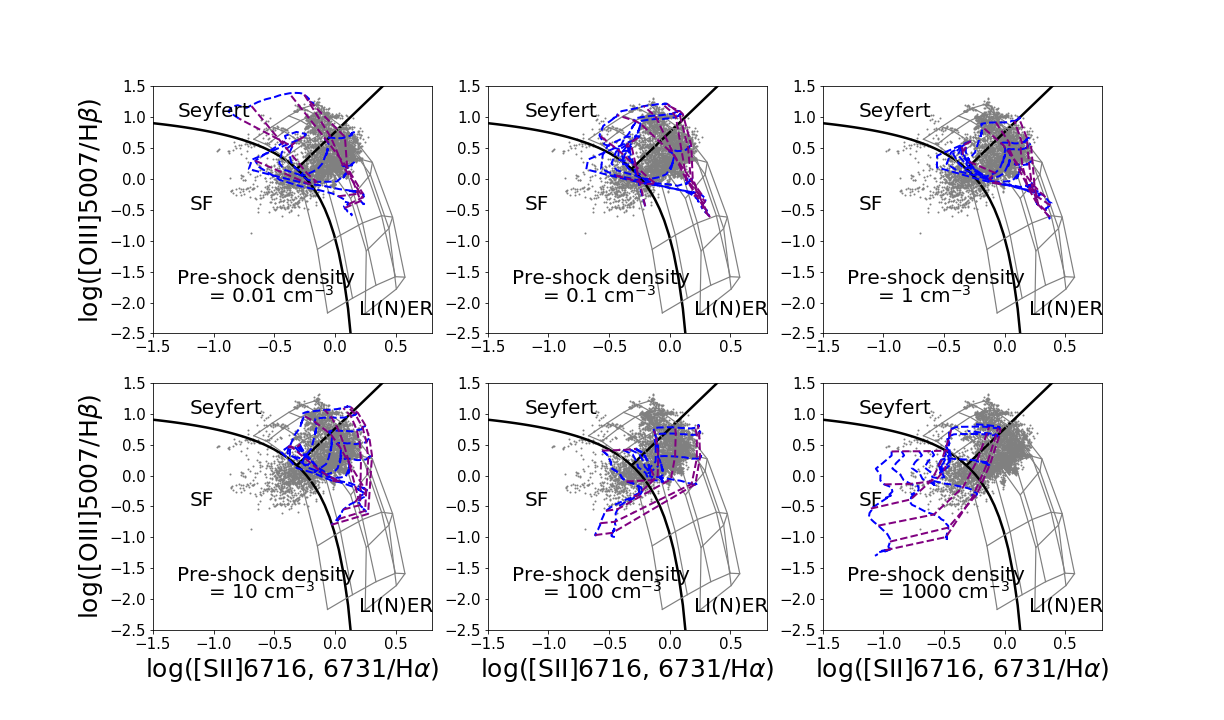}
	\caption{Shock models with different input hydrogen density (0.01 cm$^{-3}$ $\sim$ 1000 cm$^{-3}$) in the [S~II] BPT diagram. Blue dashed lines are iso-velocity lines and purple dashed lines are iso-magnetic field parameter (B/n$^{1/2}$) lines. Shock velocities vary from 100 km/s to 1000 km/s. And the magnetic field parameter varies from 10$^{-4}$ to 10 $\mu$G cm$^{3/2}$. Metallicity is assumed be solar in all cases. Grey data points are sample spaxels from MaNGA. Grey background grids are from the photoionization model with a power-law SED of which the power-law index is $-$ 1.4.}
    \label{fig:shock_den}
\end{figure*}

In Fig.~\ref{fig:shock_den}, we consider shock models with fixed metallicity (solar) but varied input hydrogen density (0.01 cm$^{-3}$ $\sim$ 1000 cm$^{-3}$). To make a fair comparison with the observational data as well as photoionization models, we calculate the output or post-shock density using the resulting line ratio [S~II]$\lambda$ 6716 / [S~II]$\lambda$ 6731, assuming a constant electron temperature of 10$^4$ K. Due to the compression by shocks, the post-shock densities are in general $\sim$ 10$^2$ times higher than the pre-shock densities. So for a shock dominated cloud with density around 100 cm$^{-3}$ to 1000 cm$^{-3}$, the corresponding pre-shock densities range from 1 cm$^{-3}$ to 10 cm$^{-3}$. From Fig.~\ref{fig:shock_den} we can see that these shock models could produce very similar line ratios as pure photoionization models. So one cannot simply rule out the possibility of shock ionization.

An interesting question is whether shock models have an upper right boundary in the BPT diagrams as do photoionization models. In the case of photoionization, we see that as metallicity increases, line ratios first increase, and then stop increasing at the boundary. After that, line ratios start to decrease as metallicity increases. Unlike photoionization grids, the shock grids we use have extra dependence on velocity as well as magnetic field strength, making its shape much more complicated. Nevertheless, we can focus on the central location of the grids for a given metallicity.

\begin{figure*}
	\centerline{\includegraphics[width=1.1\textwidth]{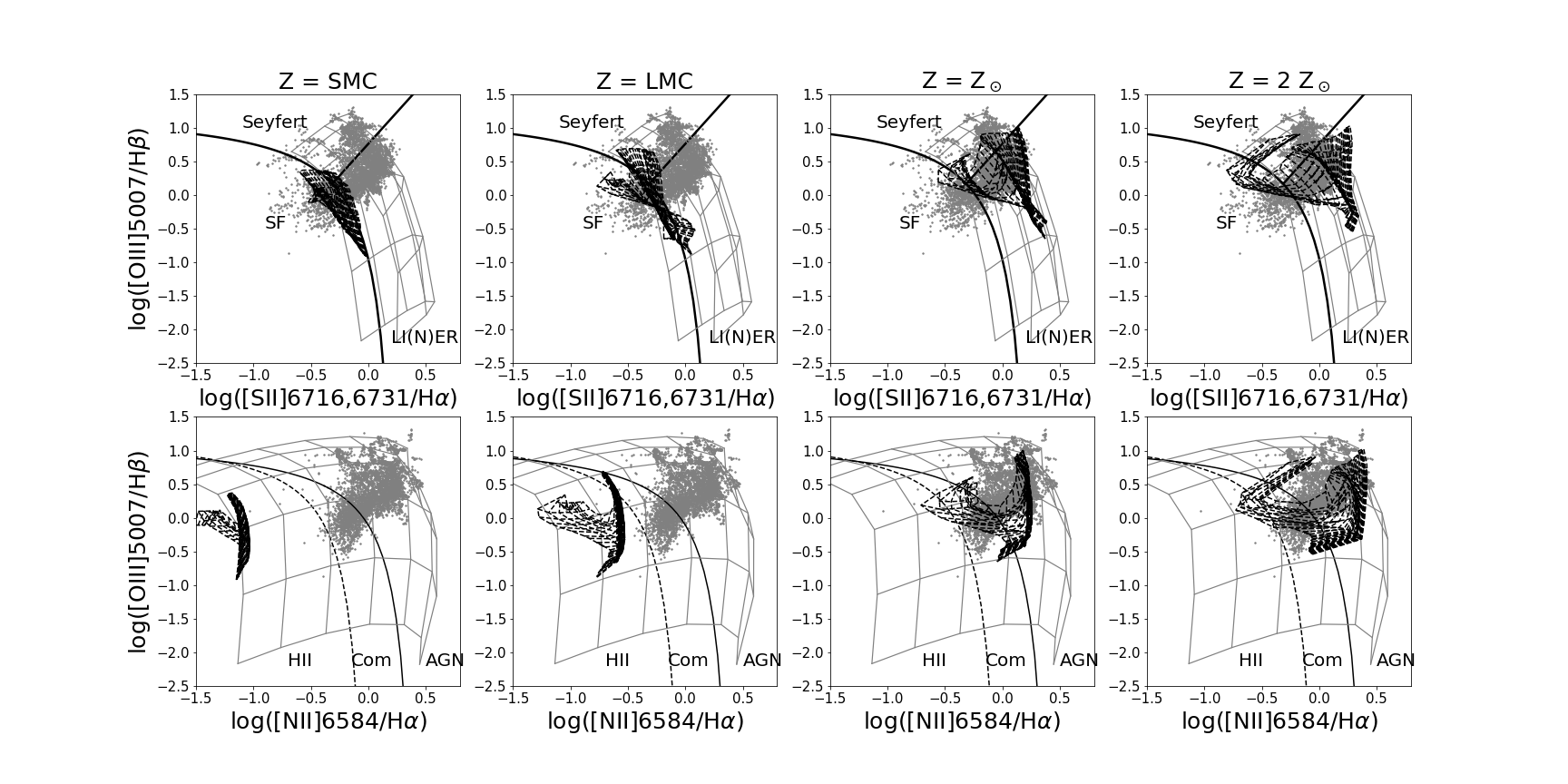}}
	\caption{Shock models with different metallicities in the [S~II] and [N~II] BPT diagrams. The pre-shock density is fixed to be 1 cm$^{-3}$. Shock velocities vary from 100 km/s to 1000 km/s. And the magnetic field parameter (B/n$^{1/2}$) varies from 10$^{-4}$ to 10 $\mu$G cm$^{3/2}$. Grey data points are sample spaxels from MaNGA. Grey background grids are from the photoionization model with a power-law SED of which the power-law index is $-$ 1.4.}
    \label{fig:shock_Z}
\end{figure*}

In Fig.~\ref{fig:shock_Z}, we compare shock models with four different metallicities. In both BPT diagrams, grids with higher metallicities are systematically shifted towards the upper-right direction. While from subsolar (SMC and LMC) to solar metallicity, the change in the location of the grid is very conspicuous, from solar to supersolar metallicity, the change is considerably smaller. So one might expect the shock grids to stop moving towards the upper right at higher metallicity, creating a boundary. But this boundary, if exists, is unlikely to be as good a match to the data as that of a photoionizing origin, as the shock model at the highest metallicity seems to exhibit a different slope from that of the observed boundary. The behavior of the shock grids indicates that the existence of an upper-right boundary in BPT diagrams could be more fundamental than the ionization mechanism, but the detailed shape of this boundary depends on the ionization mechanism. Evidently our current data disfavour a shock dominated scenario.

\subsection{Shock models in the [O~I] BPT diagram}

\begin{figure*}
	\centerline{\includegraphics[width=1.1\textwidth]{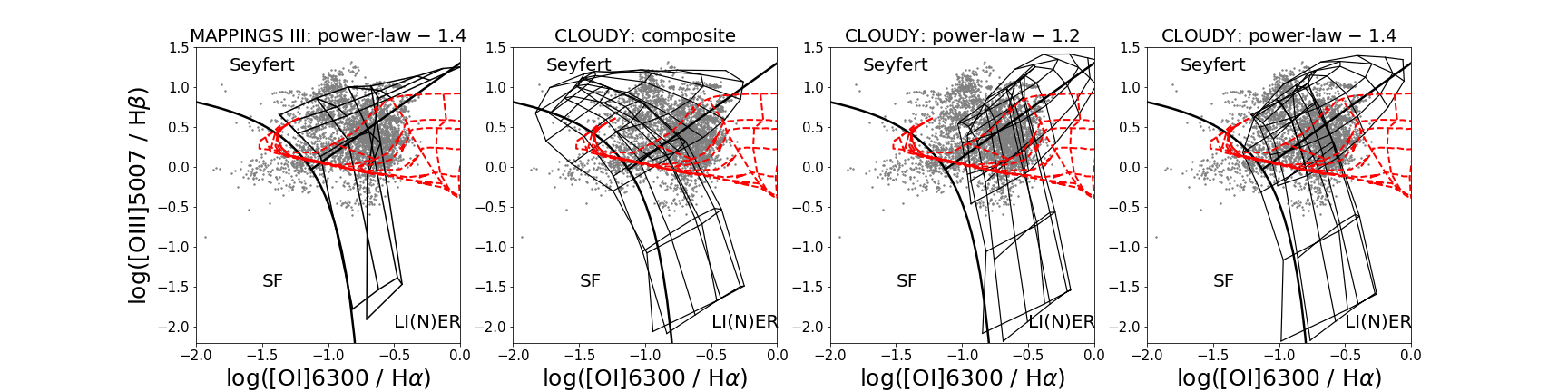}}
	\caption{[O~I] BPT diagnostic. Four sets of photoionzation models described before are shown in black. And shock model with a pre-shock density of 1 cm$^{-3}$ is shown in red.}
    \label{fig:o1_bpt}
\end{figure*}

So far we have mainly focused on the behaviour of our models in the [S~II] and [N~II] BPT diagrams, while leaving out the [O~I] BPT diagram, which is another widely used diagram in optical. Fig.\ref{fig:o1_bpt} shows our photoionzation models as well as data points in this diagram. The data points are selected from MPL-8 with the additional requirement that the S/N of [O~I]$\lambda$6300 is greater than 3. Evidently there is large discrepancy between the data and the models preferred by the other two diagrams. All three power-law models miss a number of spaxels at the top. The rightmost data points are also difficult to explain. In three of the four models, the rightmost data points can only be covered by models with ionizationparameters, q, greater than $-$ 2.0, which are ot likely given our previous analyses. As a conclusion, none of these models is fully consistent with the data points in the [O~I] BPT diagram. Increasing the electron density can shift the model grids rightwards as the [O~I] $\lambda$ 6300 line has a high critical density, making the composite SED model fit the data better. But this would bring in inconsistency with the results we get from the other two BPT diagrams as well as the calculation of electron densities of our sample. So it seems our models will inevitablely fail after taking the [O~I] diagnostic into account.

We note that there is a caveat associated with the calculation of the strength of [O~I] $\lambda$ 6300 emission line: The spatial location of [O~I] emitting region makes it sensitive to shocks and non-equilibrium heating happening at the ionization front (\citealp{2013ApJS..208...10D}). Therefore one should be cautious while trying to put this diagnostics in comparison with photoionzation models. One related question is whether this discrepancy implies the presence of shock ionization. For the shock models we have from {\tt MAPPINGS III}, the answer seems to be no, as shock models with different densities all tend to miss some spaxels at the top as photoionzation models do. In Fig.~\ref{fig:o1_bpt}, we include the shock model with a pre-shock density of 1 cm$^{-3}$. The grid is more rightwards rather than upwards compared with the photoionzation grids. As a result, the discrepancy would remain even if we include both models.

\section{Summary and conclusions}

We investigate the AGN regions in optical BPT diagrams with 4378 spaxels from the centers of 262 different galaxies in the MaNGA survey, which display a upper right boundary in the [S~II] diagnostic diagram. The boundary is not caused by selection cuts on emission lines as even the weakest line, H$\beta$, has decent SNR at the boundary. Therefore, there has to be a physical origin associated with this boundary. Similar to the case of the curved boundary in BPT diagrams for SF regions, photoionization models of AGN naturally predict such a boundary matching the shape of the observed one. The detection of this boundary in MaNGA data can put strong constrains on the shape of AGN SED.

Using photoionization models produced by the {\tt CLOUDY} code as well as {\tt MAPPINGS III} code, we examine the dependence of the position of the boundary on electron density, input SED, and secondary element abundances. We reach the following conclusions:
\begin{enumerate}
\item Our data in the [S~II] BPT diagram shows a clear boundary. The position of this boundary is insensitive to electron density and the prescription of the secondary elements. But it is very sensitive to the shape of the SED. By matching the boundary to the data, we can constrain the upper limit of the SED power-law slope to be 1.40 $\pm$ 0.05. We can also describe this boundary with a combination of two line ratios: log ([O~III] $\lambda$ 5007 / H$\beta$) = $-$ 2.1 log ([S~II] $\lambda \lambda$ 6716, 6731 / H$\alpha$) + (0.90 $\sim$ 1.00). We note that the exact location of this boundary could well depend on how one would define it given the distribution of the data. Also the variation of the intrinsic physical parameters might introduce extra uncertainties to the location of the boundary. A composite SED can produce a very similar upper-right boundary, but its behaviour at the high ionization parameter regime is significantly different from power-law models.
\item There is no clear sign of a boundary in the [N~II] BPT diagram. The coverage of the photoionization models in the [N~II] BPT diagram, however, is sensitive to the electron density, the shape of the input SED, and the nitrogen prescription. By comparing the [S II] doublet ratios of our data with the model predictions, we are able to constrain the median density of our data to be around 100 cm$^3$. With both the density and the shape of the SED fixed, models with nitrogen prescriptions yielding high N/O ratios, like the Groves04 prescription, are favored by our data.
\item Using a least square method to derive ionization parameter and metallicity with both [S~II] and [N~II] diagrams for our sample, we find good agreement in the former predictions but large discrepancy in the latter. Since the inclusion of the secondary nitrogen causes the model grids on the [N~II] BPT diagram to stretch horizontally, the metallicity derived based on this diagram tend to be super-solar, seemingly more compatible with typical NLRs in the nearby Universe. Contamination from star formation can have larger impact on the [S~II] diagnostics and contribute to the discrepancy. N/O variation due to some evolutionary factors might be another minor contributor. If the two commonly used BPT diagrams, [N~II] and [S~II] diagrams, are combined into a 3D diagnostic, some degeneracy seen in the 2D diagnostics no longer exists. A specific model would manifest itself as a 2D surface embedded in the 3D space. We see that our data exhibit a transition from the AGN surface to the SF surface, indicating the existence of SF contamination with increasing importance. This scenario is consistent with the metallicity expectation for the typical NLRs.
\item Degeneracy at high ionization parameter happens for all our models, indicating the existence of another upper boundary. However our data points do not exhibit a continuous distribution throughout the regime where q > $-$ 2.0, indicating the lack of high ionization points in our sample. Extra constraints can be applied to the detailed shape of the SED if there are more high q data points.
\item We also check the possibility of shock ionization. We find that shock models can result in similar line ratios as those from pure photoionzation models. After adding the metallicity dimension, it is possible that the shock models are also bounded by upper-right boundaries on BPT line-ratio planes. However, judging from the shape of the model grids, the shock boundaries, if exist, would not match the observed boundary as well as the photoionization models do. To further verify this point we need to explore shock models in a larger parameter space, most importantly with higher metallicity.
\item Lastly, the other commonly used optical BPT diagnostics, [O~I] diagnostic, shows inconsistency with our photoionzation models. The photoionzation calculation of [O~I] $\lambda$ 6300 might suffer from missing non-thermal heating sources. Though being a possible heating source in the ionization front, shocks alone cannot resolve the tension in the [O~I] diagnostic.
\end{enumerate}

\section*{Acknowledgements}

RY acknowledges support by NSF AST-1715898. R. R. thanks CNPq, CAPES and FAPERGS.

Funding for the Sloan Digital Sky Survey IV has been provided by the Alfred P. Sloan Foundation, the U.S. Department of Energy Office of Science, and the Participating Institutions. SDSS acknowledges support and resources from the Center for High-Performance Computing at the University of Utah. The SDSS web site is www.sdss.org.

SDSS is managed by the Astrophysical Research Consortium for the Participating Institutions of the SDSS Collaboration including the Brazilian Participation Group, the Carnegie Institution for Science, Carnegie Mellon University, the Chilean Participation Group, the French Participation Group, Harvard-Smithsonian Center for Astrophysics, Instituto de Astrof\'isica de Canarias, The Johns Hopkins University, Kavli Institute for the Physics and Mathematics of the Universe (IPMU) / University of Tokyo, the Korean Participation Group, Lawrence Berkeley National Laboratory, Leibniz Institut f\"ur Astrophysik Potsdam (AIP), Max-Planck-Institut f\"ur Astronomie (MPIA Heidelberg), Max-Planck-Institut f\"ur Astrophysik (MPA Garching), Max-Planck-Institut f\"ur Extraterrestrische Physik (MPE), National Astronomical Observatories of China, New Mexico State University, New York University, University of Notre Dame, Observat\'orio Nacional / MCTI, The Ohio State University, Pennsylvania State University, Shanghai Astronomical Observatory, United Kingdom Participation Group, Universidad Nacional Aut\'onoma de M\'exico, University of Arizona, University of Colorado Boulder, University of Oxford, University of Portsmouth, University of Utah, University of Virginia, University of Washington, University of Wisconsin, Vanderbilt University, and Yale University.




\bibliographystyle{mnras}
\bibliography{ms} 








\bsp	
\label{lastpage}
\end{document}